\documentclass[aps,prd,preprint,groupedaddress,showpacs,eqsecnum,nofootinbib,floatfix]{revtex4}
\usepackage{graphicx,epsf,subfigure,multirow,slashbox,hhline}
\graphicspath{{./plots/}}

 \newif\ifdraft
\drafttrue
\newif\ifpreprint
\preprinttrue

\def\fig#1{fig.~{\ref{#1}}}
\def\Fig#1{Fig.~{\ref{#1}}}

\def\Sect#1{Section~{\ref{#1}}}
\def\sect#1{section~{\ref{#1}}}
\def\eqn#1{eq.~(\ref{#1})}

\def\tab#1{table~{\ref{#1}}}

\newcommand{\be}{\begin{equation}}
\newcommand{\ee}{\end{equation}}
\newcommand{\ba}{\begin{eqnarray}}
\newcommand{\ea}{\end{eqnarray}}

\newcommand{\BlackHat}{{\sc BlackHat}}
\newcommand{\SHERPA}{{\sc SHERPA}}
\newcommand{\AMEGIC}{{\sc AMEGIC++}}
\newcommand{\COMIX}{{\sc COMIX}}

\newcommand{\HTj}{H_T^{\rm jet}}

\def\Zjj{$Z\,\!+\,2$}
\def\Zjjj{$Z\,\!+\,3$}
\def\Zjjjj{$Z\,\!+\,4$}

\def\Zjjx{$Z\,\!+\,2,3$}

\def\Zjn{$Z\,\!+\,n$}
\def\Wjn{$W\,\!+\,n$}
\def\Zp{$Z\,\!+\,$}
\def\Wp{$W\,\!+\,$}

\def\gjj{$\gamma\,\!+\,2$}
\def\gjjj{$\gamma\,\!+\,3$}
\def\gjjx{$\gamma\,\!+\,2,3$}
\def\gjjjj{$\gamma\,\!+\,4$}
\def\gp{$\gamma\,\!+\,$}
\def\gjm{$\gamma\,\!+\,m$}
\def\gjn{$\gamma\,\!+\,n$}
\def\gjnmo{$\gamma\,\!+\,(n\!-\!1)$}

\def\Vjj{$V\,\!+\,2$}
\def\Vjjj{$V\,\!+\,3$}
\def\Vjjjj{$V\,\!+\,4$}
\def\Vjn{$V\,\!+\,n$}

\def\METjjj{${\s E}_T\,\!+\,3$}

\def\METe{\textrm{MET}}

\def\nub{\bar\nu}
\def\qb{\bar q}

\def\Ord{{\cal O}}

\newbox\charbox
\newbox\slabox
\def\s#1{{      
        \setbox\charbox=\hbox{$#1$}
        \setbox\slabox=\hbox{$/$}
        \dimen\charbox=\ht\slabox
        \advance\dimen\charbox by -\dp\slabox
        \advance\dimen\charbox by -\ht\charbox
        \advance\dimen\charbox by \dp\charbox
        \divide\dimen\charbox by 2
        \raise-\dimen\charbox\hbox to \wd\charbox{\hss/\hss}
        \llap{$#1$}
}}
\def\root{{\sc root}}
\def\ntuple{{$n$-tuple}}
\def\pt{p_{\rm T}}
\def\ptmin{p_{\rm T}^{\rm min}}

\begin{document}
\hfuzz 10 pt

\ifpreprint
\noindent
UCLA/12/TEP/103 \hskip2.3cm
SLAC--PUB--15096
\hfill CERN--PH--TH/2011-170\\
IPPP/11/42
\hfill SB/F/407-12
\fi

\title{Missing Energy and Jets for Supersymmetry Searches}

\author{Z.~Bern${}^a$, G.~Diana${}^b$,
L.~J.~Dixon${}^{c}$, F.~Febres Cordero${}^d$, 
S.~H{\"o}che${}^c$,  H. Ita${}^{a,e}$, D.~A.~Kosower${}^b$,
D.~Ma\^{\i}tre${}^{f,g}$ and K.~J.~Ozeren${}^a$}

\affiliation{
\centerline{${}^a$Department of Physics and Astronomy, UCLA,
Los Angeles, CA 90095-1547, USA} \\ 
\centerline{${}^b$Institut de Physique Th\'eorique, CEA--Saclay,
F--91191 Gif-sur-Yvette cedex, France} \\
\centerline{${}^c$SLAC National Accelerator Laboratory,
Stanford University, Stanford, CA 94309, USA} \\
\centerline{${}^d$Departamento de F\'{\i}sica, Universidad
Sim\'on Bol\'{\i}var, Caracas 1080A, Venezuela} \\
\centerline{${}^e$Niels Bohr International Academy and Discovery Center, NBI,
 DK-2100, Copenhagen, Denmark}
\centerline{${}^f$Theory Division, Physics Department, CERN,
CH--1211 Geneva 23, Switzerland} 
\centerline{${}^g$Department of Physics, University of Durham,
Durham DH1 3LE, UK} \\
}


\begin{abstract}
We extend our investigation of backgrounds to new physics signals,
following CMS's data-driven search for supersymmetry at the LHC.  The
aim is to use different sets of cuts in \gjjj-jet production to
predict the irreducible \Zjjj-jet background (with the $Z$ boson
decaying to neutrinos) to searches with \METjjj-jet signal topologies.
We compute ratios of \Zjjj-jet to \gjjj-jet production cross sections
and kinematic distributions at next-to-leading order (NLO) in
$\alpha_s$.  We compare these ratios with those obtained using a
parton shower matched to leading-order matrix elements (ME+PS).  This
study extends our previous work~\cite{DrivingMissing} on the \Zjj-jet
to \gjj-jet ratio.  We find excellent agreement with the ratio
determined from the earlier NLO results involving two instead of three
jets, and agreement to within 10\% between the NLO and ME+PS results
for the ratios.  We also examine the possibility of large QCD
logarithms in these processes.  Ratios of \Zjn-jet to \gjn-jet cross
sections are plausibly less sensitive to such corrections than the
cross sections themselves.  Their effect on estimates of \Zjjj-jet to
\gjjj-jet ratios can be assessed experimentally by measuring the
\gjjj-jet to \gjj-jet production ratio in search regions.  We
partially address the question of potentially large electroweak
logarithms by computing the real-emission part of the electroweak
corrections to the ratio using ME+PS, and find that it is 1\% or less.
Our estimate of the remaining theoretical uncertainties in the $Z$ to
$\gamma$ ratio is in agreement with our earlier study.
\end{abstract}

\pacs{12.38.Bx, 13.85.Qk, 13.87.Ce\hspace{1cm}}

\maketitle

\renewcommand{\thefootnote}{\arabic{footnote}}
\setcounter{footnote}{0}

\section{Introduction}
\label{IntroSection}

The Large Hadron Collider has now produced more than two years of
data from high-energy collisions.
Data from the first year of running have been analyzed in a wide
variety of searches, to seek new physics beyond the Standard Model,
and to understand the underlying mechanism of electroweak symmetry-breaking.
Search topologies with large missing transverse energy (MET) accompanied
by several jets (METJ) play an important role in searches for supersymmetry
and other models of new physics containing dark matter candidates.

Events with these topologies do not automatically point to new
physics, as Standard-Model processes can give rise to similar ones.
One example is the production of a $Z$ boson in
association with jets, with the $Z$ then decaying into a pair of neutrinos
(METZJ).  METJ searches require that we understand these backgrounds.

The CMS collaboration has
used~\cite{CMSPhotonNote,CMSMET} $W$-boson and photon
production in association with jets in order to estimate the METZJ
background in setting limits on the constrained minimal supersymmetric
standard model and on simplified models of new physics~\cite{CMSSearch}.  
In such a data-driven approach, the unknown background is estimated
by combining an experimental measurement of a reference process 
(which may be the same process in a different kinematic region) with a 
theoretical factor expressing the ratio between the two processes.
This approach cancels the experimental systematics common to 
both processes, and can also reduce theoretical uncertainties.
Theoretical input is still required to estimate the ratio
and its uncertainties.  The stability of the ratio under different
theoretical approximations can be used to validate the theoretical
uncertainty.

The most obvious choice of reference process to estimate the METZJ
background would be another $Z$ decay process, where the $Z$ is again
produced in association with jets, but decays to a charged-lepton pair.
Only the $Z$ branching ratio differs from the METZJ process.  
However, the
rate for the charged-lepton process is a factor of six lower
(per lepton flavor), even before taking into account reductions
due to lepton rapidity cuts.  The low statistics in the charged-lepton
process has motivated experimenters to use other processes to estimate
METZJ rates.  The CMS collaboration has 
studied~\cite{CMSPhotonNote,CMSMET} and used~\cite{CMSSearch}
$W$-boson or photon production in association with jets for making such
estimates. The production of a $W$ in association with jets offers
an order of magnitude higher statistics than the leptonic $Z$ process;
the production of a prompt photon in association with jets,
sixteenfold higher statistics than leptonic $Z$ decays.
Photon production also avoids contamination from $t\overline t$
production.  The cuts required to suppress this
background in $W$ production enhance the photon channel's advantage.

Photon production in associated with jets has also been studied
in ref.~\cite{AskEtal} and used by the ATLAS
collaboration~\cite{ATLASPhoton} in their data-driven estimates
of the METZJ background.  Another recent study has examined the
scaling of \gp{}jets with the number of jets~\cite{EnglertEtal}.

Both \Wp{}jets and \gp{}jets production probe different combinations
of the parton distributions and different scales than \Zp{}jets
production.  The impact of these effects must be determined
theoretically.  This in turn requires a theoretical study of ratios of
photon production with respect to that of massive vector bosons.

The masslessness of the photon further requires a precise definition
of what is meant experimentally by its detection.  In the experiments,
the photons must be isolated in order to eliminate otherwise-copious hadronic
backgrounds, while overly strong isolation would lead to unwanted
vetoing due to the underlying event.
In a theoretical calculation, one must be careful to ensure that the
photon-isolation criterion is infrared- and collinear-safe.  Some QCD
radiation must be allowed near the photon.  This ensures that
corresponding cross sections and distributions can be computed
reliably in perturbation theory.  Previous theoretical studies have
used a variety of different isolation criteria, which are usually
phrased in terms of the limits on the amount of hadronic energy in a
cone surrounding the photon.  Fixed isolation cones generally limit
either the total amount of (transverse) energy in the cone, or the
hadronic energy fraction of the total (transverse) energy in the cone.
In contrast, the criterion proposed by Frixione~\cite{Frixione}
consists of a set of energy constraints that become increasingly
restrictive the closer one gets to the photon.  This latter criterion
eliminates long-distance collinear fragmentation contributions of
partons into photons.  Its attractive theoretical properties flow from
this fact.  The other cone criteria require a perturbative
factorization.  While such a factorization is available, the required
photon fragmentation functions~\cite{PhotonFragmentation}
(non-perturbative functions analogous to the parton distribution
functions) are not known very precisely.

In their study, CMS used~\cite{CMSPhotonNote,CMSMET} a fixed
hadronic-energy limit in an $R=0.4$ cone surrounding the photon.  In
our previous study~\cite{DrivingMissing}, we used a Frixione isolation
criterion.  In the same paper, we also showed that, at the large
transverse boson momenta of interest in the search, the difference
between the two isolation criteria was under 1\%, a conclusion
confirmed by a comparison to an inclusive-photon measurement by
CMS~\cite{CMSInclusivePhoton}.  Additional jets are not expected to
significantly alter this conclusion.  As part of the present study, we
have compared cross sections computed using a Frixione-type isolation
with those imposing a standard-cone isolation for both \gjj-jet and
\gjjj-jet production, and find that the two are indeed within the
expected 1\% in the regions of interest. We used a parton shower
matched to tree-level matrix elements (ME+PS) for this comparison. We
will again use the Frixione isolation criterion in the present study.

Our previous study~\cite{DrivingMissing} looked at the \Zjj-jet and
\gjj-jet production processes at next-to-leading order (NLO) accuracy
in the strong coupling $\alpha_s$.
We compared the NLO results with those computed using ME+PS.  We
provided the theoretical input needed for using the photon process to
estimate the $Z$ one, and for assessing the remaining theoretical
uncertainties in this procedure.  These results were used by the CMS
collaboration to provide the theoretical uncertainty in their search
for new physics based on topologies with large missing transverse energy
and three or more jets~\cite{CMSSearch}.

In this article, we extend our study to \Zjjj-jet and \gjjj-jet
production at NLO in $\alpha_s$.  This is the first NLO computation of 
\gjjj-jet production at a hadron collider.
In order to estimate the theoretical
uncertainties on the ratios of the two processes, we will
again compare the NLO results to the ME+PS ones. 
(The correlated variation of factorization and renormalization scales
in the numerator and denominator of these ratios produces only small
shifts in the ratios, which are likely to underestimate the
uncertainties substantially.)  We study these processes both with the
selection cuts used by CMS~\cite{CMSSearch} and studied for \gjj-jet
and \Zjj-jet production in ref.~\cite{DrivingMissing}, and also with a
set of tighter selection cuts.  As we shall see, our results are
consistent with our previous study, and indeed, the NLO ratios
computed using \Vjjj-jet production are remarkably similar to those
computed using \Vjj-jet production, where $V$ stands for both $Z$ and
$\gamma$.

The comparison of \Vjjj-jet to \Vjj-jet production reveals potentially
significant QCD logarithms, related to ratios of large scalar transverse
energy and MET requirements to small minimum jet transverse momenta,
which we examine.
Liu {\it et al.}~\cite{PetrielloQCDLogs} have recently resummed 
a different class of logarithms (of threshold type) in \Vjj-jet production
processes. These are very similar to threshold logarithms previously
resummed in pure QCD~\cite{OtherThresholdLogs}, and related to
threshold logs resummed in top-quark production~\cite{TopResum}.
However, we are not aware of a comprehensive study of other
large logarithms that may arise in such \Vjn-jet production processes.
At very large energies, virtual electroweak corrections are potentially
significant, due to Sudakov double
logarithms~\cite{ElectroweakLogsMaina,ElectroweakLogsKuhn}.
As in ref.~\cite{DrivingMissing}, we do not include these virtual effects.
However, we have used the \SHERPA{} parton-shower code to estimate the
effects of radiating an additional real electroweak gauge boson, which
decays hadronically.  While this is not a detailed study, it suggests 
that the real-radiation effects are much smaller than those induced by
virtual corrections~\cite{ElectroweakLogsMaina,ElectroweakLogsKuhn}.

We employ the same software tools as in our previous studies of \Wjn-jet
and \Zjn-jet
production~\cite{W3j,W4j,BlackHatZ3jet,Wpolarization,Z4j}:
the \BlackHat{} library~\cite{BlackHatI,BlackHatII} implementing on-shell
methods numerically, along with \AMEGIC{}~\cite{Amegic} within the
\SHERPA{}~\cite{Sherpa} framework, to perform the leading-order (LO)
and NLO calculations.  We also use the \SHERPA{} framework to obtain
the ME+PS results.  The public version of \SHERPA{} does not properly
treat the $Z$ and $\gamma$ cases on an equal footing, causing a bias as
the number of jets increases. To obtain sensible predictions for the
\Zjjj-jet to \gjjj-jet ratio we have modified it somewhat.

This paper is organized as follows.  In \sect{CalculationSection}
we outline our calculation.  \Sect{CutsSection} discusses the
various cuts we use.  In \sect{BasicPredictionSection} we present the
total cross sections for \Zjj-jet, \Zjjj-jet, \gjj-jet and \gjjj-jet
production for the different regions. In \sect{JetRatioSection} we
examine jet production ratios (the ratio of \Vjn~jets to
$V\,\!+\,(n-1)$ jets) in various regions of phase space.
In section~\ref{ZgammaStabilitySection} we present the ratios
of \Zjjj-jet to \gjjj-jet rates for cross sections, along with the
corresponding ratios of \Zjj-jet to \gjj-jet rates, and selected
distributions. In \sect{UncertaintySection}, we compare NLO QCD to
ME+PS predictions and obtain an estimate of remaining theoretical
uncertainties in the $Z$ to $\gamma$ ratio.  We give our conclusions
and outlook in \sect{ConclusionSection}.  In the appendix we describe
how we modified \SHERPA{} so that $Z$ bosons and photons are
treated on an equal footing.

\begin{figure}[t]
\begin{center}
\subfigure[]{\includegraphics[clip,scale=0.4]{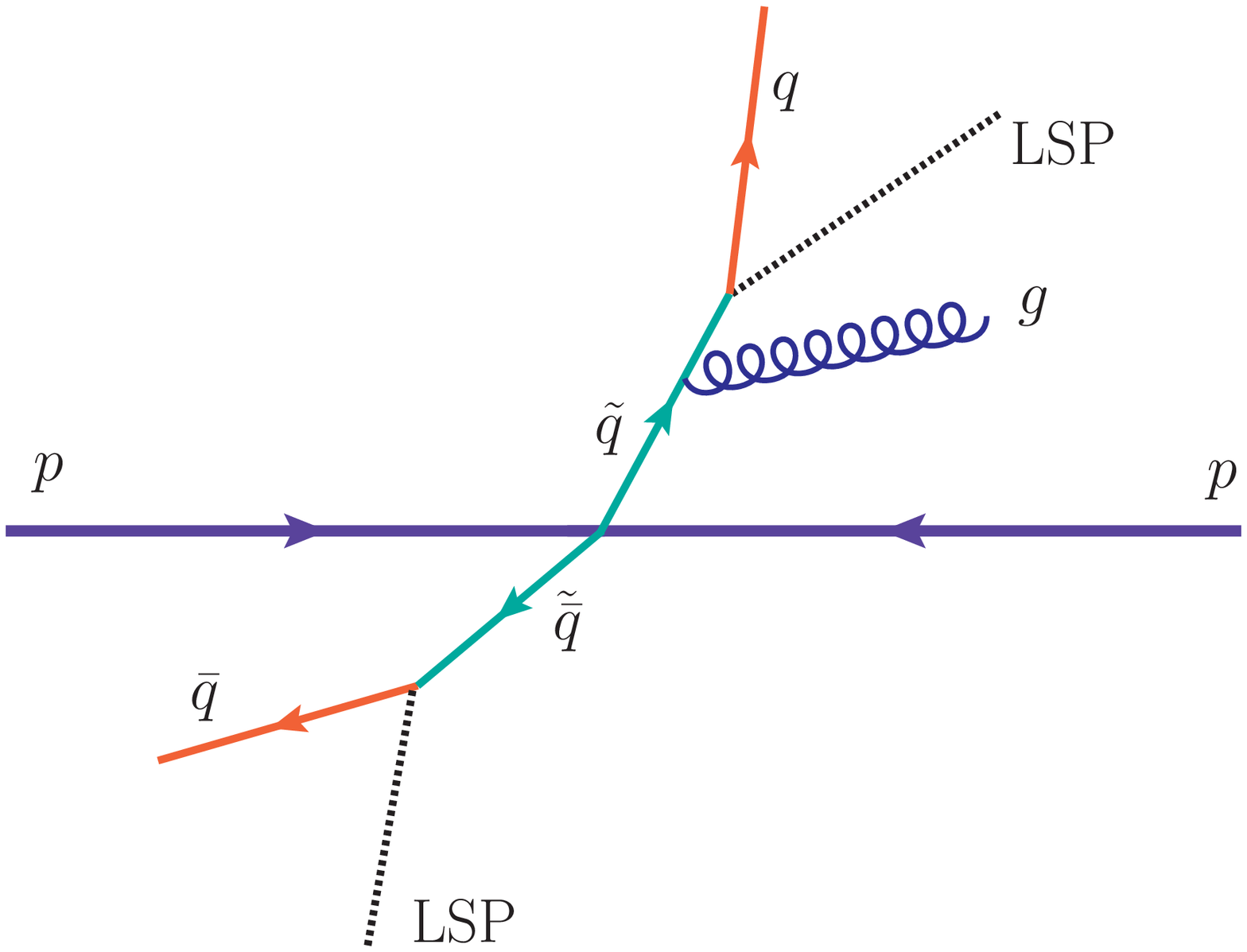}}
\subfigure[]{\includegraphics[clip,scale=0.4]{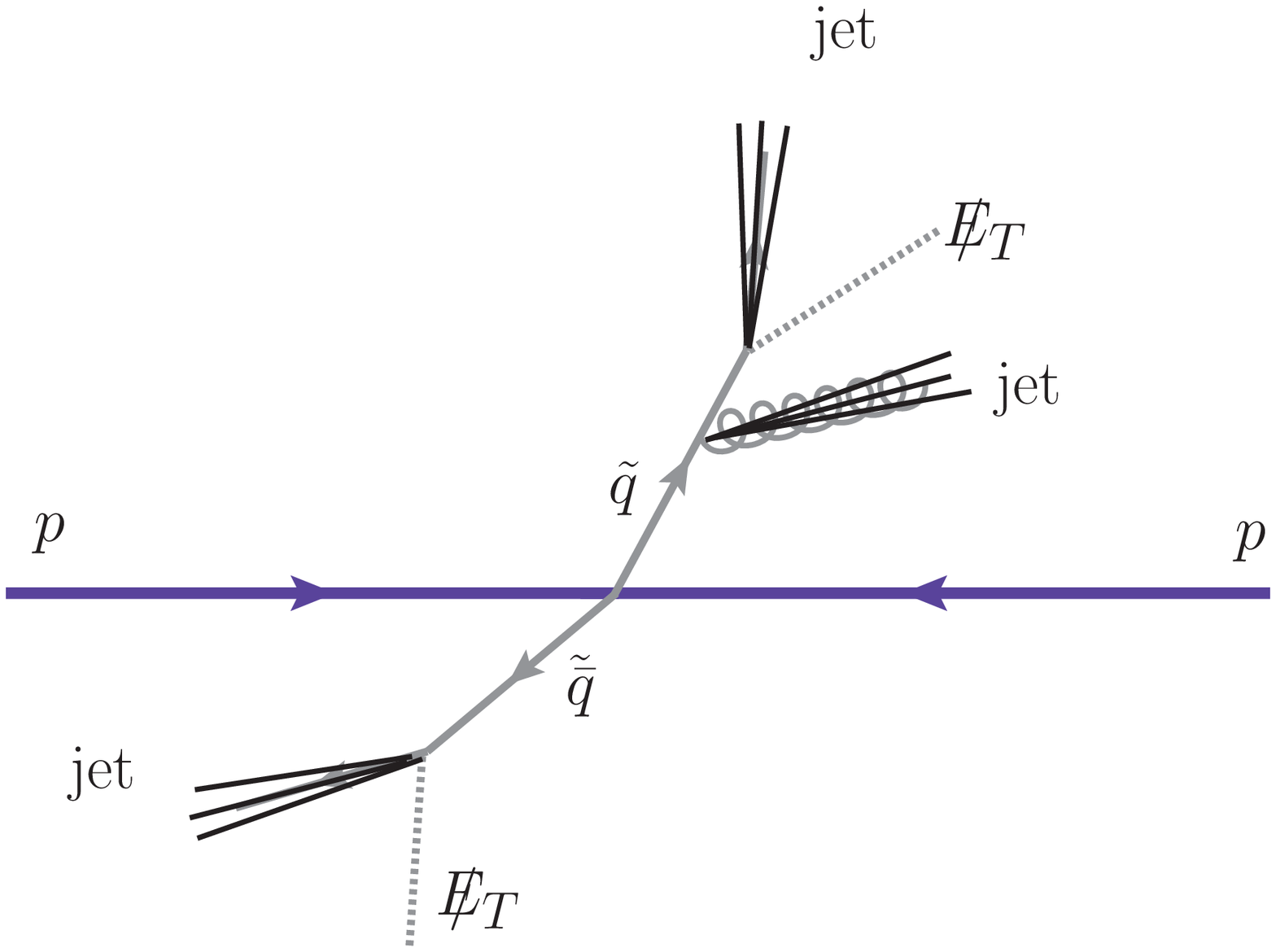}}\\
\end{center}
\caption{Squark pair production illustrates a new-physics process with
the signature of three jets plus MET.  Here each squark decays to a quark
and the lightest neutralino; the escaping neutralinos generate the 
missing transverse energy. }
\label{fig:NewPhysicsExample}
\end{figure}

\section{The Calculation}
\label{CalculationSection}

We compute the cross sections at NLO in fixed-order perturbation theory,
following the same basic organization as in previous
studies~\cite{BlackHatZ3jet,W3j,W4j,DrivingMissing,Z4j}.  We combine
several contributions: the LO term; virtual corrections from the
interference of tree-level and one-loop amplitudes; the real-emission
corrections with dipole subtraction~\cite{CS} terms; and the singular
phase-space integrals of the dipole terms.

\begin{figure}[t]
\begin{center}
 \subfigure[]{\label{fig:Zdiagram}\includegraphics[clip,scale=0.45]{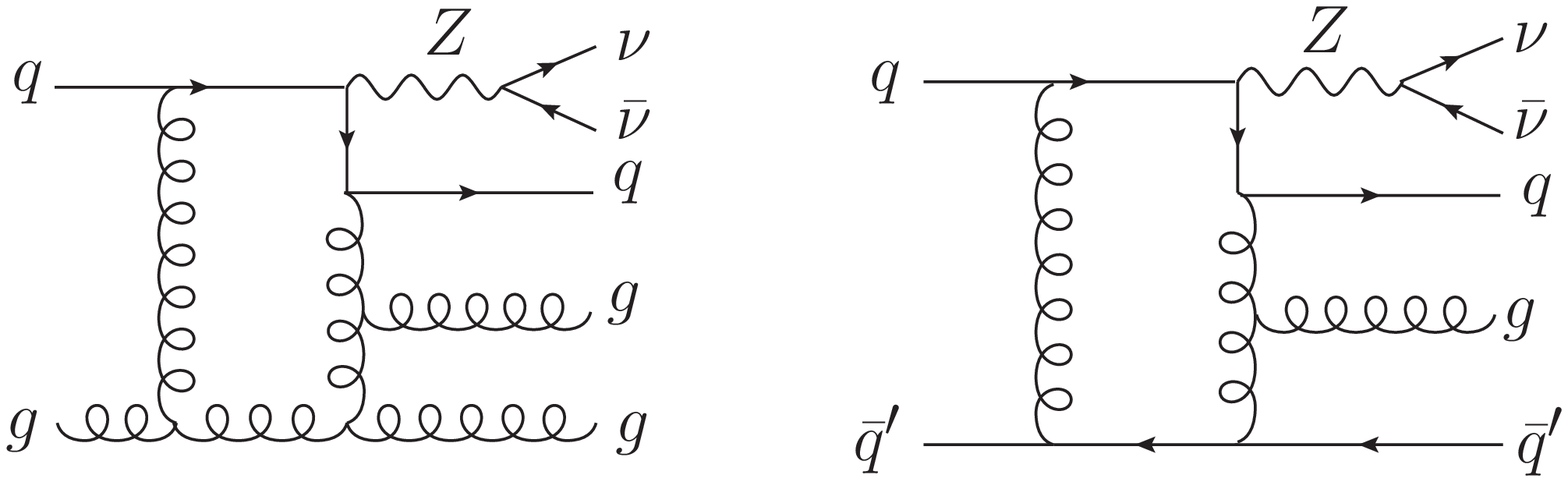}}\\
 \subfigure[]{\label{fig:Gamdiagrams}\includegraphics[clip,scale=0.45]{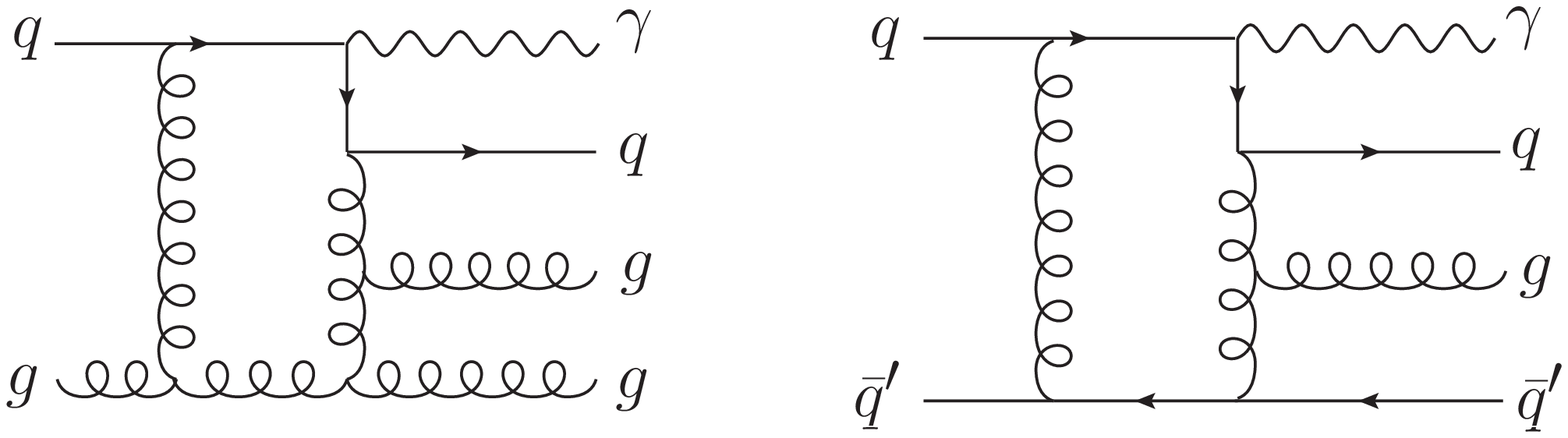}}
\end{center}
\caption{Sample virtual diagrams needed for (a) 
$pp \rightarrow Z(\rightarrow \nu\nub)+3$-jet production and
for (b) $pp \rightarrow \gamma+3$-jet production.}
\label{fig:VirtualExamples}
\end{figure}

We evaluate the required one-loop amplitudes using the \BlackHat{}
program library~\cite{BlackHatI}, which implements on-shell
methods numerically.  For the processes we are studying,
we need the one-loop corrections to the following partonic processes,
\begin{eqnarray}
 && q \qb g g g\rightarrow Z(\rightarrow \nu\nub)\ \hbox{or}\ \gamma\,, 
\nonumber \\
 && q \qb q' \qb' g\rightarrow Z(\rightarrow \nu\nub)\ \hbox{or}\ \gamma\,,
\label{VirtualProcesses}
\end{eqnarray}
where three of the five partons are crossed into the final state, and the
$Z$ decay to neutrinos is folded in.  We compute
both distinct- and identical-quark flavor
subprocesses for the second partonic process.
We exhibit sample diagrams for
these processes in \fig{fig:VirtualExamples}, illustrating
the similarity of the $Z$ and $\gamma$ cases.

For both $Z$ and $\gamma$ processes, the \BlackHat{} code
library~\cite{BlackHatI,BlackHatII} computes the required primitive
amplitudes using a numerical implementation of on-shell methods.  The
photonic primitive amplitudes are obtained directly, rather than as
sums over permutations of color-ordered primitive amplitudes for purely
colored partons, as was done for the photonic amplitudes in our previous
study~\cite{DrivingMissing}.  We omit the process $gg\to ggg\gamma$ as
it contributes to \gjjj-jet production only at $\Ord(\alpha_S^5)$, two orders
higher than the LO processes in \eqn{VirtualProcesses}.
At the large values of parton $x$ of interest here,
the gluon luminosity is not large enough to compensate for the
additional powers of $\alpha_s$.  
 As in refs.~\cite{BlackHatZ3jet,Z4j},
we drop small vector and axial loop contributions, along with the small
effects of top quarks.  However, all subleading color contributions
are included. (See ref.~\cite{ItaOzeren} for a general method for doing so.)

The NLO result also requires real-emission corrections to the LO
process, which arise from tree-level amplitudes with one additional
parton.  We use the~\AMEGIC{} code~\cite{Amegic}, included in the
\SHERPA{} framework, to compute these contributions, along with the
Catani-Seymour dipole subtraction terms~\cite{CS} and their integrals
over phase space.  We have previously
validated~\cite{W3j,BlackHatZ3jet} the \BlackHat+\SHERPA{} framework
for $W,Z + (n\le2)$ jets against the MCFM code~\cite{MCFM}.

In our study, we wish to vary the renormalization and factorization scales, 
and also to make use of parton distribution function (PDF) error sets 
to estimate associated uncertainties.  To do so efficiently,
we organize all contributions into sums of terms (in an automated
way),
where each term contains a simple function we wish to vary (for example,
a logarithm of the renormalization scale) multiplied by a
numerical coefficient independent of such variation.  
We calculate these coefficients in one run, and store them for re-use. 
For each event we generate, we record the momenta for all partons
along with the coefficients of the various
scale- or PDF-dependent functions.
  We store this information in \root{}-format \ntuple{}
files~\cite{ROOT}.  The availability of these intermediate results in
a standard format makes it straightforward for us to evaluate
cross sections and distributions for different scales and PDF error
sets.  We can also furnish theoretical predictions to
experimental collaborations by handing over \ntuple{} files.
The experiments can modify the cuts applied, or compute additional
distributions~\cite{ATLASZJets,ATLASWJets,PureJets}. 

We use the parton shower implemented in \SHERPA{} to compute a
parton-shower prediction matched to tree-level matrix elements
(ME+PS), also known as matrix-element-plus-truncated-shower. 
The ME+PS event samples are produced following ref.~\cite{HoechePhoton},
using the \COMIX{} matrix-element generator~\cite{Comix}.  (We
expect the $Z$ to $\gamma$ ratios to be insensitive to
hadronization effects; accordingly, to allow a cleaner comparison to
the fixed-order NLO results, we do not include hadronization effects
but present the ME+PS results at parton level.)  However, as we
explain in greater detail in the appendix, we do not
use a standard public version of \SHERPA{}.  Instead we modify version~1.3.1,
in order to ensure that low-scale radiation in \Zjn-jet and \gjn-jet
production is treated on the same footing.

We work to leading order in the electroweak coupling. The $Z$-boson
couplings we use are given in ref.~\cite{BlackHatZ3jet}.  The
$\nu\nub$ invariant mass is distributed in a relativistic Breit-Wigner
of width $\Gamma_Z = 2.49$~GeV about the $Z$ boson mass of
$91.1876$~GeV. These values, along with $\alpha_{\rm QED}(M_Z) =
1/128.802$ and $\sin^2\theta_W = 0.230$, lead to a branching ratio for
the neutrino mode in $Z$ decay of ${\rm Br}(Z\to\nu\nub) = 0.2007$.
We use MSTW2008 LO and NLO parton distribution functions, with the QCD
coupling $\alpha_s$ chosen appropriately in each case. Our results are
for an LHC center-of-mass energy of $7$ TeV.  As explained in
ref.~\cite{DrivingMissing} (see also refs.~\cite{Alpha}), we use the
zero-momentum-squared value, $\alpha_{\rm EM}(0) = 1/137.036$, for the
electromagnetic coupling in the photon amplitudes.

Photon measurements make use of an isolation criterion.  Experimental
collaborations typically use a weighted isolation criterion (see {\it
  e.g.} ref.~\cite{CMSPhotonNote}), imposing a limit on the hadronic
energy fraction in a cone around the photon, or simply on the total
hadronic energy in the cone.  The theoretical version of this
criterion requires the use of nonperturbative photon fragmentation
functions. Frixione~\cite{Frixione} proposed a modified isolation
requirement which suppresses the collinear region of the phase space
and thereby eliminates the need for a fragmentation-function
contribution.  We follow this proposal, requiring that each parton $i$
within a distance $R_{i\gamma}$ of the photon obey
\begin{equation} 
\label{iso}
\sum\limits_i E_{iT}\, \Theta \left(\delta - R_{i\gamma}\right)
   \leq {\mathcal{H}} (\delta) \,,
\end{equation}
for all $\delta \leq \delta_0$, in a cone of fixed half-angle $\delta_0$
around the photon axis. The restricting function ${\mathcal{H}} (\delta)$ is
chosen such that it vanishes as $\delta \rightarrow 0$ and thus suppresses
collinear configurations, but allows soft radiation arbitrarily close to the
photon. We adopt 
\be
{\mathcal{H}} (\delta) = E_{T}^\gamma \, \epsilon 
\left( \frac{ 1 - \cos \delta }{1 - \cos \delta_0 } \right)^n \, , 
\label{photoniso}
\ee
where $E_T^\gamma$ is the photon transverse energy.  

As in our previous study of \Zjj-jet and \gjj-jet
production~\cite{DrivingMissing}, we will use the Frixione cone, with
$\epsilon = 0.025$, $\delta_0 = 0.3$ and $n = 2$.  We studied the
sensitivity to these parameters in ref.~\cite{DrivingMissing}, and
found it to be weak\footnote{We follow our previous study in the
  details of the jet algorithm: to obtain the cross section for \gjm{}
  jets, we apply the jet-finding algorithm to all partons except the
  photon, whether inside the isolation cone or not.  We insist that
  there be $m$ jets lying outside the photon isolation cone that pass
  the jet rapidity and minimum-$\pt$ cuts.}.  We also compared the
predictions using these parameters to predictions made using
standard-cone isolation of the isolated prompt-photon
spectrum~\cite{CMSInclusivePhoton} measured by CMS.  We found the
differences between the Frixione and standard-cone isolation
prescriptions to be relatively small, and less than 1\% in the
large-$\pt^\gamma$ region that is our primary interest.  We concluded
that it was reasonable to use the Frixione isolation to model the
$Z$ to $\gamma$ ratio in association with two jets for CMS's analysis,
and that the same conclusion should hold for the $Z$ to $\gamma$ ratio in
association with three jets.  We have now used a ME+PS calculation
to compare directly \gjjj-jet production using Frixione isolation
to a standard cone isolation mimicking CMS's criterion.
This parton-shower calculation
effectively includes only the perturbative contributions to the
photon fragmentation function.  Nonetheless, we again find that
the two isolation criteria give very similar results in the high-$\pt$
region, agreeing to within 1\%, which buttresses our previous
conclusion~\cite{DrivingMissing}.

\section{Control and Search Regions}
\label{CutsSection}

\def\firstJ{{\rm 1}^{\it st}{\, \rm jet}}
\def\secondJ{{\rm 2}^{\it nd}{\, \rm jet}}

Our focus in this paper is on using distributions measured for
inclusive \gjjj-jet production to predict similar distributions
assembled from inclusive missing~$E_T$\,+\,3-jet events.  We focus on
two control regions and five search regions suggested to us by the CMS
collaboration. One of these control regions and two of the search
regions were used to set limits on supersymmetry and other new physics
models, based on data collected in 2010~\cite{CMSSearch}.
The other control and search regions, relevant for the much larger
2011 data set, apply harder cuts which push out further on the
tails of the underlying distributions.  The different search regions
are intended to be relevant in different regions of the parameter space
of supersymmetric extensions to the Standard Model. We have
generated a set of \ntuple{}s, implementing the weakest of all
the cuts we list below while 
generating events.  The full cuts are applied during the
analysis of \ntuple{} files.

We follow CMS and use the anti-$k_T$ jet algorithm~\cite{antikT}
with clustering parameter $R = 0.5$ throughout, where
$R = \sqrt{(\Delta y)^2+(\Delta\phi)^2}$ is the usual distance
measure in terms of rapidity difference $\Delta y$ and azimuthal angle
difference $\Delta\phi$.  Jets are ordered in $\pt$.

The CMS cuts make use of a special definition of the total transverse
energy, which we label $\HTj$. It is the scalar sum of the transverse energies
of all jets with $\pt > 50$~GeV and pseudorapidity $|\eta| < 2.5$.  We also
define\footnote{MET stands for `Missing Transverse Energy', but in fact
  denotes the missing transverse momentum.}
a vector MET, as the negative of the sum of the transverse momenta of all
jets with $\pt > 30$~GeV and $|\eta|<5$.
Each region that we consider
is distinguished by a different set of cuts on the quantities $\HTj$
and $|\textrm{MET}|$:
\begin{description}
    \item[Set 1:] {$\HTj > 300$ GeV}, {$|\textrm{MET}| > 250$ GeV};
    \item[Set 2:] {$\HTj > 500$ GeV}, {$|\textrm{MET}| > 150$ GeV};
    \item[Set 3:] {$\HTj > 300$ GeV}, {$|\textrm{MET}| > 150$ GeV};
    \item[Set 4:] {$\HTj > 350$ GeV}, {$|\textrm{MET}| > 200$ GeV};
    \item[Set 5:] {$\HTj > 500$ GeV}, {$|\textrm{MET}| > 350$ GeV};
    \item[Set 6:] {$\HTj > 800$ GeV}, {$|\textrm{MET}| > 200$ GeV};
    \item[Set 7:] {$\HTj > 800$ GeV}, {$|\textrm{MET}| > 500$ GeV}.
\end{description}
The cuts in Sets 1--3 are the same as those used by the CMS
collaboration~\cite{CMSSearch} and were also used in our previous
study of \Zjj-jet and \gjj-jet production.  Sets 4--7 impose harder
(tighter) cuts, appropriate for searches with larger data sets.
In addition to computing the \Zjjj-jet and \gjjj-jet cross sections,
we repeat our previous 2-jet study, extending it to the new
kinematic sets.

For all sets we require three jets with at least $50$ GeV of
transverse momentum and absolute pseudorapidity of at most
$2.5$. These jets are called
`tagging jets'.  The azimuthal separation between the two leading
tagging jets and the MET vector is required to satisfy
$\Delta\phi(\textrm{jet}_i,\textrm{MET}) > 0.5$, $i=1,2$.  We require that the
jet with the third-highest $\pt$ also be separated from the MET
vector, $\Delta\phi(\textrm{jet}_3,\textrm{MET}) > 0.3$.  Additional
jets beyond the third are not subject to such a constraint.
(When repeating the \Vjj-jet study, obviously
we require only two tagging jets, and the last constraint does not
apply.)

In addition to the above cuts, for the \gjjx{}-jet cross sections
only, we impose photon isolation according to the
Frixione~\cite{Frixione} prescription, with parameters $\epsilon =
0.025$, $\delta_0 = 0.3$ and $n = 2$ in \eqn{photoniso}.
We also follow CMS and require a minimum $R$-space
separation between the MET vector and each tagging jet of 0.4.
The photon is required to have $|\eta| < 2.5$.
We impose no explicit minimum $\pt$ on the vector boson, although
the MET cuts make it very likely to have large $\pt$.

The Set 1 cuts can be roughly characterized as the low-$\HTj$ /
high-MET region, whereas Set 2 is the converse, high-$\HTj$ /
low-MET. The reason for studying these two sets is that different
SUSY production mechanisms are expected to lead to signals in
different regions. Broadly speaking, Set 1 is geared towards catching
direct squark decays, while Set 2 is designed for cascades with a
$W$ boson and a softer lightest supersymmetric particle (LSP). Set 3,
which is inclusive of both the others, is a control region.  Set 4
is again a control region, and is inclusive of the regions
covered by Sets~5, 6, and 7.  These sets push further into the tail
of distributions, and are designed to search for heavier superpartners.

Our fixed-order results depend on the renormalization scale $\mu_R$ and
factorization scale $\mu_F$. These scales are unphysical, and hence
physical cross sections should be independent of them; but 
a dependence on them necessarily
appears when the perturbative series is truncated at a finite order.  For
fixed-order predictions, it is customary to estimate the uncertainty
arising from omission of higher-order terms by varying these scales
around some central value. The size of the resulting band is a useful
diagnostic for those situations where fixed-order perturbation theory
breaks down. The central value should be a typical hard scale
in the process, to minimize the impact of potentially large
logarithms. We choose the dynamical scale $\mu = \mu_R = \mu_F = H_T'/2$
for this central value, where $H_T'$ is defined as the scalar
transverse energy sum,
\begin{equation}
H_T' = \sum_i E_T^i + E_T(Z,\gamma) \,,
\end{equation}
with $i$ running over the partons and $E_T(V) \equiv \sqrt{ M_V^2+\pt^2}$. 
We evaluate cross sections
at five values of the common renormalization and factorization scale:
 $\mu/2, \mu/\sqrt2, \mu,
\sqrt2\mu, 2\mu$.  As we will discuss below, this procedure is
expected to greatly underestimate uncertainties when applied to a
ratio of cross sections with similar QCD properties. 

\section{Basic LHC Predictions}
\label{BasicPredictionSection}

\begin{table}[t]
  \begin{center}
\begin{tabular}{|c|c||l|l|}
\hline
Set & Prediction & \multicolumn{1}{c|}{\Zjjj-jet} & 
\multicolumn{1}{c|}{\gjjj-jet}  \\
\hhline{|=|=#=|=|} 
\multirow{3}{*}{1}
& LO & ~~$0.200(0.001)^{+0.105}_{-0.064}$~~ & ~~$0.856(0.002)^{+0.446}_{-0.273}$~~ \\
\cline{2-4}
& ME+PS & ~~$0.157(0.001)$~~ & ~~$0.772(0.01)$~~ \\
\cline{2-4}
& NLO & ~~$0.186(0.002)^{+0.007}_{-0.023}$~~ & ~~$0.830(0.01)^{+0.049}_{-0.109}$~~ \\
\hhline{|=|=#=|=|} 
\multirow{3}{*}{2}
& LO & ~~$0.1790(0.0005)^{+0.095}_{-0.058}$~~ & ~~$0.913(0.002)^{+0.479}_{-0.292}$~~ \\
\cline{2-4}
& ME+PS & ~~$0.160(0.002)$~~ & ~~$0.844(0.01)$~~ \\
\cline{2-4}
& NLO & ~~$0.170(0.002)^{+0.007}_{-0.022}$~~ & ~~$0.868(0.01)^{+0.041}_{-0.109}$~~ \\
\hhline{|=|=#=|=|} 
\multirow{3}{*}{3}
& LO & ~~$0.664(0.001)^{+0.346}_{-0.211}$~~ & ~~$3.46(0.01)^{+1.780}_{-1.090}$~~ \\
\cline{2-4}
& ME+PS & ~~$0.533(0.01)$~~ & ~~$3.09(0.04)$~~ \\
\cline{2-4}
& NLO & ~~$0.622(0.005)^{+0.022}_{-0.077}$~~ & ~~$3.25(0.03)^{+0.119}_{-0.396}$~~ \\
\hhline{|=|=#=|=|} 
\multirow{3}{*}{4}
& LO & ~~$0.291(0.001)^{+0.153}_{-0.093}$~~ & ~~$1.354(0.003)^{+0.704}_{-0.431}$~~ \\
\cline{2-4}
& ME+PS & ~~$0.235(0.002)$~~ & ~~$1.21(0.01)$~~ \\
\cline{2-4}
& NLO & ~~$0.270(0.003)^{+0.009}_{-0.033}$~~ & ~~$1.29(0.01)^{+0.065}_{-0.166}$~~ \\
\hhline{|=|=#=|=|} 
\multirow{3}{*}{5}
& LO & ~~$0.0341(0.0001)^{+0.0182}_{-0.0111}$~~ & ~~$0.1392(0.0004)^{+0.074}_{-0.045}$~~ \\
\cline{2-4}
& ME+PS & ~~$0.0284(0.0003)$~~ & ~~$0.124(0.002)$~~ \\
\cline{2-4}
& NLO & ~~$0.0319(0.001)^{+0.0017}_{-0.0044}$~~ & ~~$0.132(0.001)^{+0.006}_{-0.017}$~~ \\
\hhline{|=|=#=|=|} 
\multirow{3}{*}{6}
& LO & ~~$0.0185(0.0001)^{+0.0099}_{-0.0060}$~~ & ~~$0.0839(0.0004)^{+0.0450}_{-0.0273}$~~ \\
\cline{2-4}
& ME+PS & ~~$0.0173(0.0002)$~~ & ~~$0.0795(0.001)$~~ \\
\cline{2-4}
& NLO & ~~$0.0181(0.0003)^{+0.0015}_{-0.0026}$~~ & ~~$0.0814(0.001)^{+0.0058}_{-0.0114}$~~ \\
\hhline{|=|=#=|=|} 
\multirow{3}{*}{7}
& LO & ~~$0.00275(0.00002)^{+0.00152}_{-0.00091}$~~ & ~~$0.0107(0.0001)^{+0.0059}_{-0.0035}$~~ \\
\cline{2-4}
& ME+PS & ~~$0.00245(0.00003)$~~ & ~~$0.0100(0.0002)$~~ \\
\cline{2-4}
& NLO & ~~$0.00267(0.0001)^{+0.00027}_{-0.00043}$~~ & ~~$0.0105(0.0002)^{+0.0008}_{-0.0016}$~~ \\
\hline
\end{tabular}
\caption{Cross sections in pb for $Z$ and $\gamma$ production in
  association with three jets for the cuts of Sets 1--7 given in
  \sect{CutsSection}.  The numbers in parentheses are Monte Carlo
  statistical errors, while the upper and lower limits represent scale
  dependence. 
  \label{AllSets3jXS} }
  \end{center}
\end{table}

In this section we present total cross sections for the seven
control and search regions defined in the previous section.
We present results for \gjjj-jet and \Zjjj-jet production at the LHC for
$\sqrt{s} = 7$ TeV.  We also update our previous results~\cite{DrivingMissing}
for \gjj-jet and \Zjj-jet production for the cuts of Sets 1--3,
and extend them to Sets 4--7.  In the \Zjj-jet and \Zjjj-jet studies, we fold
in the decay of the $Z$ boson into neutrinos, which in turn give rise to
missing transverse momentum.
The branching ratio for the $Z$ decay to neutrinos is largely responsible
for the \gjjj-jet cross section being about a
factor of four to five larger than for $Z(\to\nu\nub)\,\!+\,3$ jets.
The value of this ratio is the primary underlying motivation for our
study, and is clearly visible in our tables and
figures.
In later sections, we will study various ratios constructed from the
numbers presented here.

In \tab{AllSets3jXS}, we display the total cross section for the
different sets of cuts detailed in \sect{CutsSection}.  For each set,
we show three different theoretical predictions for the \Zjjj-jet and
\gjjj-jet cross sections in sequence: LO; ME+PS; and NLO.  The final
states in the fixed-order cases (LO and NLO) consist of the vector boson
with the three tagging jets, and possibly an extra jet at NLO.  In the
parton-shower case, the final state can contain many jets, although
virtual corrections are not taken into account.  The ME+PS calculation
is computed using \SHERPA, modified from the public version 1.3.1 as
explained in the appendix.
The LO fixed-order predictions are the least reliable of the three
and are shown only for reference purposes.

\begin{table}[t]
  \begin{center}
\begin{tabular}{|c|c||l|l|}
\hline
Set & Prediction & \multicolumn{1}{c|}{\Zjj-jet} & 
\multicolumn{1}{c|}{\gjj-jet}\\
\hhline{|=|=#=|=|} 
\multirow{3}{*}{1}
& LO & ~~$0.512(0.001)^{+0.188}_{-0.128}$~~ & ~~$2.050(0.002)^{+0.745}_{-0.508}$~~ \\
\cline{2-4}
& ME+PS & ~~$0.432(0.002)$~~ & ~~$1.93(0.02)$~~ \\
\cline{2-4}
& NLO & ~~$0.546(0.002)^{+0.023}_{-0.050}$~~ & ~~$2.40(0.01)^{+0.204}_{-0.267}$~~ \\
\hhline{|=|=#=|=|} 
\multirow{3}{*}{2}
& LO & ~~$0.2002(0.0003)^{+0.075}_{-0.051}$~~ & ~~$0.933(0.001)^{+0.346}_{-0.235}$~~ \\
\cline{2-4}
& ME+PS & ~~$0.236(0.002)$~~ & ~~$1.14(0.01)$~~ \\
\cline{2-4}
& NLO & ~~$0.272(0.002)^{+0.038}_{-0.038}$~~ & ~~$1.35(0.01)^{+0.215}_{-0.201}$~~ \\
\hhline{|=|=#=|=|} 
\multirow{3}{*}{3}
& LO & ~~$1.234(0.001)^{+0.445}_{-0.304}$~~ & ~~$5.780(0.005)^{+2.050}_{-1.410}$~~ \\
\cline{2-4}
& ME+PS & ~~$1.16(0.01)$~~ & ~~$6.12(0.04)$~~ \\
\cline{2-4}
& NLO & ~~$1.445(0.005)^{+0.116}_{-0.156}$~~ & ~~$7.50(0.02)^{+0.894}_{-0.944}$~~ \\
\hhline{|=|=#=|=|} 
\multirow{3}{*}{4}
& LO & ~~$0.509(0.001)^{+0.188}_{-0.127}$~~ & ~~$2.179(0.002)^{+0.794}_{-0.540}$~~ \\
\cline{2-4}
& ME+PS & ~~$0.486(0.002)$~~ & ~~$2.28(0.01)$~~ \\
\cline{2-4}
& NLO & ~~$0.600(0.003)^{+0.051}_{-0.067}$~~ & ~~$2.80(0.01)^{+0.333}_{-0.357}$~~ \\
\hhline{|=|=#=|=|} 
\multirow{3}{*}{5}
& LO & ~~$0.0561(0.0001)^{+0.0217}_{-0.0146}$~~ & ~~$0.2179(0.0004)^{+0.084}_{-0.056}$~~ \\
\cline{2-4}
& ME+PS & ~~$0.0544(0.0003)$~~ & ~~$0.228(0.003)$~~ \\
\cline{2-4}
& NLO & ~~$0.0664(0.0004)^{+0.0061}_{-0.0079}$~~ & ~~$0.270(0.001)^{+0.030}_{-0.034}$~~ \\
\hhline{|=|=#=|=|} 
\multirow{3}{*}{6}
& LO & ~~$0.0170(0.0001)^{+0.0066}_{-0.0044}$~~ & ~~$0.0731(0.0002)^{+0.0285}_{-0.0191}$~~ \\
\cline{2-4}
& ME+PS & ~~$0.0220(0.0002)$~~ & ~~$0.0946(0.001)$~~ \\
\cline{2-4}
& NLO & ~~$0.0245(0.0002)^{+0.0042}_{-0.0039}$~~ & ~~$0.109(0.001)^{+0.020}_{-0.018}$~~ \\
\hhline{|=|=#=|=|} 
\multirow{3}{*}{7}
& LO & ~~$0.00330(0.00001)^{+0.00136}_{-0.00089}$~~ & ~~$0.01274(0.00005)^{+0.0052}_{-0.0034}$~~ \\
\cline{2-4}
& ME+PS & ~~$0.00377(0.00003)$~~ & ~~$0.0144(0.0001)$~~ \\
\cline{2-4}
& NLO & ~~$0.00433(0.00004)^{+0.00062}_{-0.00064}$~~ & ~~$0.0170(0.0001)^{+0.0026}_{-0.0026}$~~ \\
\hline
\end{tabular}
\caption{Cross sections in pb
  for $Z$ and $\gamma$ production in association with
  two jets
  for the cuts of Sets 1--7 given in \sect{CutsSection}.
  The numbers in
  parentheses are Monte Carlo statistical errors, while the upper and
  lower limits represent scale dependence. 
  \label{AllSets2jXS} }
  \end{center}
\end{table}

In all sets, the corrections from LO to NLO are modest
at the central value of $\mu$.  The LO results
are up to 9\% larger. This is in sharp contrast to the \Zjj-jet and
\gjj-jet results presented in ref.~\cite{DrivingMissing}, and
recomputed here in \tab{AllSets2jXS}, where the LO results are up to
34\% lower.  The larger corrections in \Vjj-jet production are
expected, because the LO kinematics for \Vjj-jet production are more
constrained than those for \Vjjj-jet production, and they are relaxed
considerably when going to NLO or ME+PS kinematics.

As in our earlier study, the
ME+PS and NLO results again do not agree well for the $Z$
and $\gamma$ cross sections separately.  We do not expect the LO
or ME+PS calculations to get the overall normalization correct.
We will discuss the $Z$ to $\gamma$ ratios of these results
in~\sect{ZgammaStabilitySection}.

The results for \Zjj-jet and \gjj-jet production for Sets 1--3
differ slightly from our previous results~\cite{DrivingMissing}.
In the earlier study, we used six-flavor running of $\alpha_s$,
whereas here we use five-flavor running in order to be
consistent with the parton distributions used.  (Neither approach is completely
theoretically consistent, because of the absence of a generated
top-quark distribution.)  With five-flavor running, the beta function
is larger in magnitude, and hence $\alpha_s(\mu)$ decreases more rapidly
above $M_Z$ than with six-flavor running.  Thus the \Zjj-jet production
cross sections here are expected to be a bit smaller than those in 
ref.~\cite{DrivingMissing}. A naive estimate, based on the change in the
value of $\alpha_s$, suggests that 2\% is the right magnitude of the
difference, and this is indeed what we see in practice.


\section{Jet Production Ratios}
\label{JetRatioSection}

\begin{figure}[t]
  \begin{center}
    \includegraphics[clip,scale=0.85]{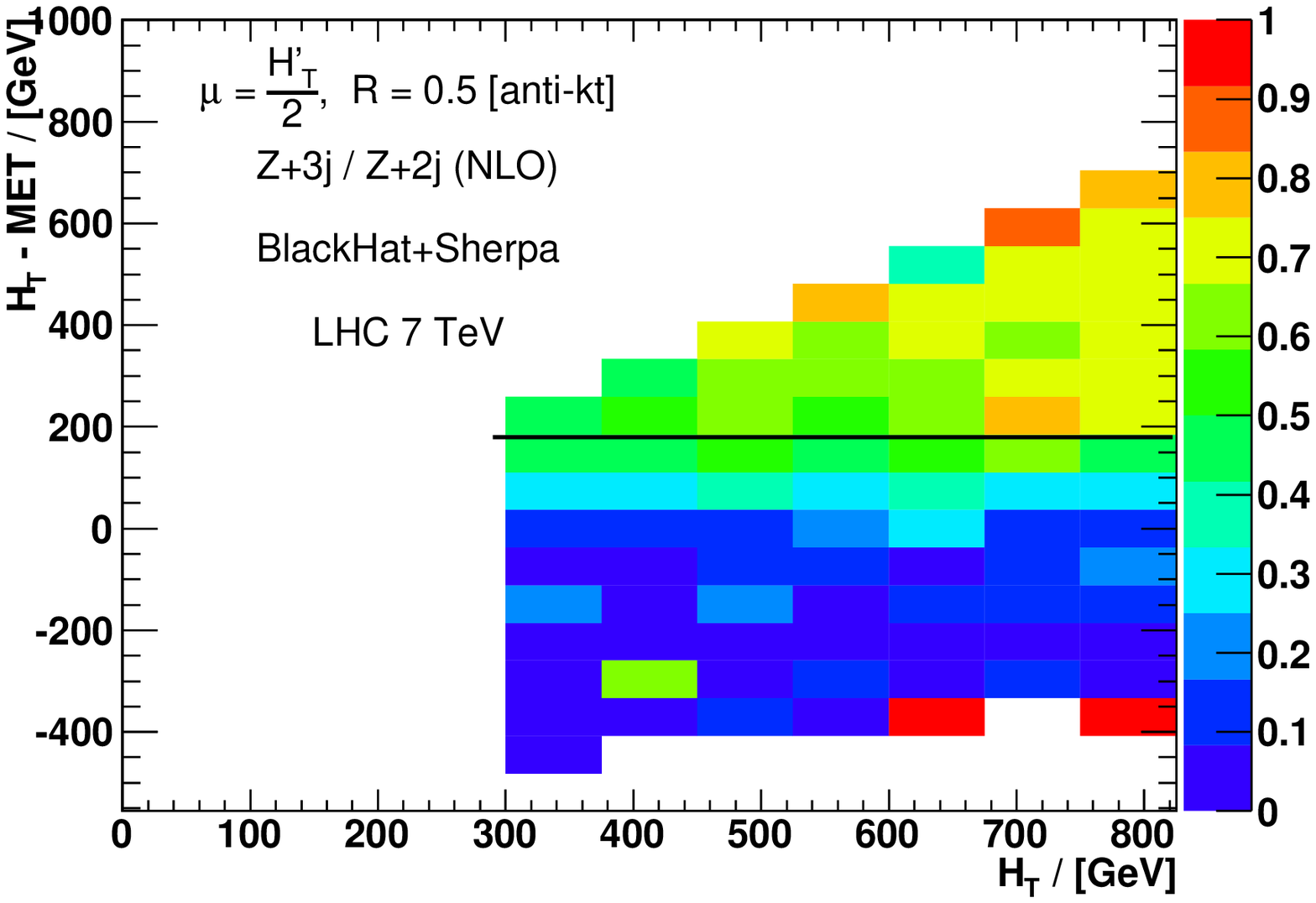}
  \end{center}
  \caption{\baselineskip 14pt%
The \Zjjj-jet to \Zjj-jet ratio as a function of $\HTj$
and $\HTj-|\textrm{MET}|$.  The solid line shows
where the ratio is roughly 0.5. }
  \label{fig:Z3-Z2-density}
\end{figure}

\def\dent{~~~~~~}
\begin{table}[t]
  \begin{center}
\begin{tabular}{|c|c||l|l|}
\hline
Set & Prediction & \multicolumn{1}{c|}{~~\Zjjj-jet/\Zjj-jet~~} & 
\multicolumn{1}{c|}{~~\gjjj-jet/\gjj-jet~~} \\
\hhline{|=|=#=|=|} 
\multirow{3}{*}{1}
& LO & \dent$0.390(0.001)$~~ & \dent$0.418(0.001)$~~ \\
\cline{2-4}
& ME+PS & \dent$0.364(0.004)$~~ & \dent$0.399(0.01)$~~ \\
\cline{2-4}
& NLO & \dent$0.340(0.005)$~~ & \dent$0.346(0.003)$~~ \\
\hhline{|=|=#=|=|} 
\multirow{3}{*}{2}
& LO & \dent$0.894(0.003)$~~ & \dent$0.978(0.003)$~~ \\
\cline{2-4}
& ME+PS & \dent$0.680(0.01)$~~ & \dent$0.742(0.01)$~~ \\
\cline{2-4}
& NLO & \dent$0.625(0.01)$~~ & \dent$0.643(0.01)$~~ \\
\hhline{|=|=#=|=|} 
\multirow{3}{*}{3}
& LO & \dent$0.538(0.001)$~~ & \dent$0.599(0.001)$~~ \\
\cline{2-4}
& ME+PS & \dent$0.458(0.01)$~~ & \dent$0.504(0.01)$~~ \\
\cline{2-4}
& NLO & \dent$0.431(0.004)$~~ & \dent$0.433(0.004)$~~ \\
\hhline{|=|=#=|=|} 
\multirow{3}{*}{4}
& LO & \dent$0.572(0.001)$~~ & \dent$0.621(0.001)$~~ \\
\cline{2-4}
& ME+PS & \dent$0.483(0.005)$~~ & \dent$0.532(0.01)$~~ \\
\cline{2-4}
& NLO & \dent$0.450(0.01)$~~ & \dent$0.462(0.004)$~~ \\
\hhline{|=|=#=|=|} 
\multirow{3}{*}{5}
& LO & \dent$0.608(0.003)$~~ & \dent$0.639(0.002)$~~ \\
\cline{2-4}
& ME+PS & \dent$0.523(0.01)$~~ & \dent$0.544(0.01)$~~ \\
\cline{2-4}
& NLO & \dent$0.481(0.01)$~~ & \dent$0.490(0.01)$~~ \\
\hhline{|=|=#=|=|} 
\multirow{3}{*}{6}
& LO & \dent$1.09(0.01)$~~ & \dent$1.15(0.01)$~~ \\
\cline{2-4}
& ME+PS & \dent$0.789(0.01)$~~ & \dent$0.840(0.01)$~~ \\
\cline{2-4}
& NLO & \dent$0.735(0.01)$~~ & \dent$0.744(0.01)$~~ \\
\hhline{|=|=#=|=|} 
\multirow{3}{*}{7}
& LO & \dent$0.833(0.01)$~~ & \dent$0.840(0.01)$~~ \\
\cline{2-4}
& ME+PS & \dent$0.649(0.01)$~~ & \dent$0.694(0.01)$~~ \\
\cline{2-4}
& NLO & \dent$0.617(0.02)$~~ & \dent$0.622(0.01)$~~ \\
\hline
\end{tabular}
\caption{Ratios of cross sections for \Zjjj-jet to \Zjj-jet and
  \gjjj-jet to \gjj-jet and their ratio for the cuts of Sets 1--7
  given in \sect{CutsSection}.  The numbers in parentheses are Monte
  Carlo statistical errors.
  \label{AllSets32ratioXS} }
  \end{center}
\end{table}

The cuts presented in the previous section are quite different from typical
cuts used to measure Standard-Model processes.  They push the kinematic
configurations far out onto tails of corresponding distributions.  This
introduces large ratios of scales, for example the ratio between 
$\HTj$ and the minimum transverse momentum of a jet, $\ptmin$.
Such large ratios can give rise to large logarithms.  If sufficiently
large, they may spoil the applicability of perturbation theory. 

Before examining the results for the various sets of cuts presented in
the previous section, let us explore the presence of large corrections
which may be due to such logarithms.  Our past
studies~\cite{BlackHatZ3jet} have shown that jet production ratios are
convenient tools for this purpose.  We begin by examining the
\Zjjj-jet to \Zjj-jet and \gjjj-jet to \gjj-jet ratios.

In \tab{AllSets32ratioXS}, we show the values of these ratios for the
different cut sets at LO, for the ME+PS calculation, and at NLO.  At
LO, the transverse momentum of the leading jet must be at least half
the $\HTj$ in \Zjj-jet or \gjj-jet events; at NLO, this kinematic
constraint is relaxed by real radiation.  Accordingly, the LO
distribution suffers large corrections in some regions.  While an
analogous kinematic relaxation does occur in \Zjjj-jet and \gjjj-jet
production when going from LO to NLO, the effect is much smaller.  As
a result, the \Zjjj-jet to \Zjj-jet and \gjjj-jet to \gjj-jet ratios
suffer large NLO corrections for some ranges of $\HTj$; the LO values
are typically 25--50\% larger than the NLO ones.

We expect the NLO ratios to be more reliable, and focus on them.  We
show the \Zjjj-jet to \Zjj-jet ratio in \fig{fig:Z3-Z2-density} as a
function of $\HTj$ and $\HTj-|\textrm{MET}|$.  The ratio depends much
more strongly on $\HTj-|\textrm{MET}|$ than on $\HTj$ alone.  At LO,
$\HTj-|\textrm{MET}|$ is necessarily positive, but the presence of
additional radiation at NLO allows it to become negative.

When looser cuts typical of Standard-Model measurements are applied,
with jet $\pt > 25$~GeV, the \Zjjj-jet to \Zjj-jet ratio is around
0.23~\cite{Z4j}.  For jet $\pt > 30$~GeV, the ratio drops slightly to
about 0.21, in agreement with the LHC
data~\cite{ATLASZJets,CMSWZJets}.  In \fig{fig:Z3-Z2-density}, the
darkest (blue) regions correspond to $\HTj$ and $|\METe|$ values for
which the ratio is at most moderately enhanced above 0.21; lighter
(green) areas, where the enhancement is noticeable, with a ratio
around 0.5; and the lightest (yellow) areas, where the enhancement is
significant, and the ratio approaches unity.  At small or negative
$\HTj-|\METe|$, the ratio is only moderately enhanced.  The
enhancement grows with growing $\HTj-|\METe|$, and is roughly
independent of $\HTj$ alone, when holding $\HTj-|\METe|$ fixed.  (The
red squares at the bottom of the plot are Monte Carlo statistical
fluctuations due to very small cross sections in this region.)  The
figure also shows a line corresponding roughly to a ratio of 0.5. We
take this value to be the boundary between a region where the
perturbative predictions are reliable, and a region where we cannot be
as confident in them.  Though our choice of boundary is arbitrary, it
is motivated by the good agreement between theory and experiment for
3-jet to 2-jet ratios without vector bosons, up to a value of
0.5~\cite{PureJets}.  We do not display the corresponding plot for the
case where the $Z$ boson is replaced by a photon; it is similar. It is
plausible that, even in the region where the \Zjjj-jet to \Zjj-jet
ratio is above 0.5, large QCD enhancements will be independent of the
parton distributions, and therefore will cancel in ratios such as the
\Zjjj-jet to \gjjj-jet ratio.  However, we have no proof of the
completeness of this cancellation, and prefer to be conservative and
acknowledge a lower reliability for the perturbative prediction in the
region where the \Vjjj-jet to \Vjj-jet ratio is larger than 0.5.


\begin{table}[t]
  \begin{center}
\begin{tabular}{|c|c|c|c|}
\hline
Set & \multicolumn{1}{c|}{~~\Zjjjj-jet/\Zjjj-jet~~} & 
\multicolumn{1}{c|}{~~\gjjjj-jet/\gjjj-jet~~} \\
\hline
1 & $0.233(0.004)$ & $0.254(0.003)$ \\
\hline
2 & $0.451(0.010)$ & $0.491(0.006)$ \\
\hline
3 & $0.260(0.006)$ & $0.280(0.003)$ \\
\hline
4 & $0.287(0.010)$ & $0.314(0.004)$ \\
\hline
5 & $0.341(0.010)$ & $0.358(0.005)$ \\
\hline
6 & $0.616(0.020)$ & $0.617(0.009)$ \\
\hline
7 & $0.464(0.030)$ & $0.465(0.010)$ \\
\hline
\end{tabular}
\caption{Ratios of LO cross sections for \Zjjjj-jet to \Zjjj-jet 
 and \gjjjj-jet to \gjjj-jet
  for the cuts of Sets 1--7 given in \sect{CutsSection}.
  The numbers in
  parentheses are Monte Carlo statistical errors.
  \label{AllSets43ratioXS} }
  \end{center}
\end{table}

For Set 1, the NLO ratios are about 0.35; for Sets~3, 4, and~5,
the ratios are larger but below 0.5; for the remaining Sets~(2, 6,
and~7), the ratios are bigger than 0.5 though still below 1.
All these ratios are noticeably larger than the inclusive ratio with
standard QCD measurement cuts~\cite{Z4j,ATLASZJets}, 
for some sets larger by a substantial factor.  
For Set 1, it is not large enough to spoil the applicability of
perturbation theory, and it is reasonable to assume this extends
to Sets~3, 4, and~5.  This assessment is reinforced by an
examination of the \Zjjjj-jet to \Zjjj-jet and \gjjjj-jet to \gjjj-jet
ratios at LO, shown in \tab{AllSets43ratioXS}.  
For Sets~2, 6, and~7, where the ratios are larger
than 0.45, as discussed above one should be more cautious.  
Accordingly, our confidence in our uncertainty estimates for these
sets is weaker, and would be weaker still for other, harder, search cuts.
One cannot, of course,
determine a precise value at which perturbation theory breaks down,
but rather ranges where an investigation of potential logarithms and their 
resummation may be required.
We note that the ME+PS predictions for the \Vjjj-jet to \Vjj-jet
production ratios, shown for 
comparison in \tab{AllSets32ratioXS}, are typically between the NLO and
LO results, and lie closer in value to NLO than to LO.

The \gjn-jet to \gjnmo-jet ratios discussed above can of course
be measured experimentally.  It would be interesting to do so
for the search cuts listed above.  This measurement,
in regions where one may question the applicability of 
unresummed QCD perturbation theory, could serve to increase
our confidence in the use of purely perturbative tools to 
estimate QCD corrections to \Zjn-jet to \gjn-jet ratios; or alternatively, 
to assess what additional corrections may be needed.

\section{Stability of the $Z$ to $\gamma$ Ratio}
\label{ZgammaStabilitySection}

\def\dent{~~~~~~}
\begin{table}[t]
  \begin{center}
\begin{tabular}{|c|c||l|l|l|}
\hline
Set & Prediction & 
\multicolumn{1}{c|}{~~\Zjjj-jet/\gjjj-jet~~} & 
\multicolumn{1}{c|}{~~\Zjj-jet/\gjj-jet~~} & 
\multicolumn{1}{c|}{ratio} \\
\hhline{|=|=#=|=|=|} 
\multirow{3}{*}{1}
& LO & \dent$0.233(0.001)$~~ & \dent$0.2499(0.0004)$~~ & ~~$0.933(0.004)$~~ \\
\cline{2-5}
& ME+PS & \dent$0.204(0.003)$~~ & \dent$0.224(0.002)$~~ & ~~$0.913(0.02)$~~ \\
\cline{2-5}
& NLO & \dent$0.224(0.004)$~~ & \dent$0.227(0.001)$~~ & ~~$0.984(0.02)$~~ \\
\hhline{|=|=#=|=|=|} 
\multirow{3}{*}{2}
& LO & \dent$0.196(0.001)$~~ & \dent$0.2145(0.0005)$~~ & ~~$0.914(0.004)$~~ \\
\cline{2-5}
& ME+PS & \dent$0.190(0.003)$~~ & \dent$0.207(0.002)$~~ & ~~$0.916(0.02)$~~ \\
\cline{2-5}
& NLO & \dent$0.196(0.003)$~~ & \dent$0.201(0.001)$~~ & ~~$0.973(0.02)$~~ \\
\hhline{|=|=#=|=|=|} 
\multirow{3}{*}{3}
& LO & \dent$0.192(0.001)$~~ & \dent$0.2134(0.0003)$~~ & ~~$0.899(0.003)$~~ \\
\cline{2-5}
& ME+PS & \dent$0.173(0.003)$~~ & \dent$0.190(0.001)$~~ & ~~$0.908(0.02)$~~ \\
\cline{2-5}
& NLO & \dent$0.191(0.002)$~~ & \dent$0.193(0.001)$~~ & ~~$0.994(0.01)$~~ \\
\hhline{|=|=#=|=|=|} 
\multirow{3}{*}{4}
& LO & \dent$0.215(0.001)$~~ & \dent$0.2336(0.0003)$~~ & ~~$0.922(0.003)$~~ \\
\cline{2-5}
& ME+PS & \dent$0.194(0.003)$~~ & \dent$0.213(0.002)$~~ & ~~$0.908(0.01)$~~ \\
\cline{2-5}
& NLO & \dent$0.209(0.003)$~~ & \dent$0.215(0.001)$~~ & ~~$0.973(0.01)$~~ \\
\hhline{|=|=#=|=|=|} 
\multirow{3}{*}{5}
& LO & \dent$0.245(0.001)$~~ & \dent$0.257(0.001)$~~ & ~~$0.952(0.01)$~~ \\
\cline{2-5}
& ME+PS & \dent$0.230(0.004)$~~ & \dent$0.239(0.004)$~~ & ~~$0.961(0.02)$~~ \\
\cline{2-5}
& NLO & \dent$0.242(0.01)$~~ & \dent$0.246(0.002)$~~ & ~~$0.981(0.02)$~~ \\
\hhline{|=|=#=|=|=|} 
\multirow{3}{*}{6}
& LO & \dent$0.220(0.002)$~~ & \dent$0.232(0.001)$~~ & ~~$0.948(0.01)$~~ \\
\cline{2-5}
& ME+PS & \dent$0.218(0.004)$~~ & \dent$0.232(0.003)$~~ & ~~$0.940(0.02)$~~ \\
\cline{2-5}
& NLO & \dent$0.222(0.01)$~~ & \dent$0.224(0.002)$~~ & ~~$0.988(0.03)$~~ \\
\hhline{|=|=#=|=|=|} 
\multirow{3}{*}{7}
& LO & \dent$0.257(0.003)$~~ & \dent$0.259(0.001)$~~ & ~~$0.992(0.01)$~~ \\
\cline{2-5}
& ME+PS & \dent$0.244(0.01)$~~ & \dent$0.261(0.003)$~~ & ~~$0.935(0.02)$~~ \\
\cline{2-5}
& NLO & \dent$0.254(0.01)$~~ & \dent$0.255(0.003)$~~ & ~~$0.993(0.03)$~~ \\
\hline
\end{tabular}
\caption{Ratios of cross sections for \Zjjj-jet to \gjjj-jet 
 and \Zjj-jet to \gjj-jet and their ratio
  for the cuts of Sets 1--7 given in \sect{CutsSection}.
  The numbers in
  parentheses are Monte Carlo statistical errors.
  \label{AllSetsZgratio} }
  \end{center}
\end{table}

We turn next to a discussion of the target ratio, that between
\Zjn-jet and \gjn-jet production.  In \tab{AllSetsZgratio}, we show
the predicted ratio for each of the seven regions, based both
on \Vjjj-jet production and on \Vjj-jet production.  The fixed-order
predictions in the latter column are in good agreement with our
previous study~\cite{DrivingMissing}.  The last column shows the ratio
of these predictions.

In the ratios, the LO scale variation cancels nearly
completely, if we vary the scale identically in the \Zjjj-jet
and \gjjj-jet predictions, and correspondingly in the \Zjj-jet and
\gjj-jet predictions.  In the NLO case the scale variation is a bit
larger but also very small.  Both scale variations lead to changes in
the ratio of less than $0.5\%$.  This nearly complete cancellation of
the scale variation cannot be interpreted as a small theoretical
uncertainty.  We will instead use the closeness of the NLO and ME+PS
ratios as an indication that the theoretical uncertainties for the
individual cross sections do indeed largely cancel in the ratio.  We
account separately for the uncertainties due to the parton
distributions.  The $Z$ to $\gamma$ ratios depend mostly on
the parton distributions through the $d(x)/u(x)$ ratio, and for regions
of $x$ where this quantity is relatively well measured.  Hence
the parton distribution uncertainties are small.

It is interesting that the NLO predictions for the $Z$ to $\gamma$ ratios in
all sets are quite stable under the addition of a jet.  That is, the
predictions based on the ratio of \Zjj-jet to \gjj-jet production are
quite similar to those based on the \Zjjj-jet to \gjjj-jet ratio.  For the
older sets (Sets 1--3), the predictions agree within 3\%, and even for
the newer sets (Sets 4--7), with harder cuts, the predictions agree to
within 5\%.  The LO predictions differ by up to 10\%, with the ME+PS
results mostly in between in percentage difference.  The NLO results
should be used as the central values to which experimental
measurements are compared.

\begin{figure}[t]
  \begin{center}
    \includegraphics[clip,scale=0.40]{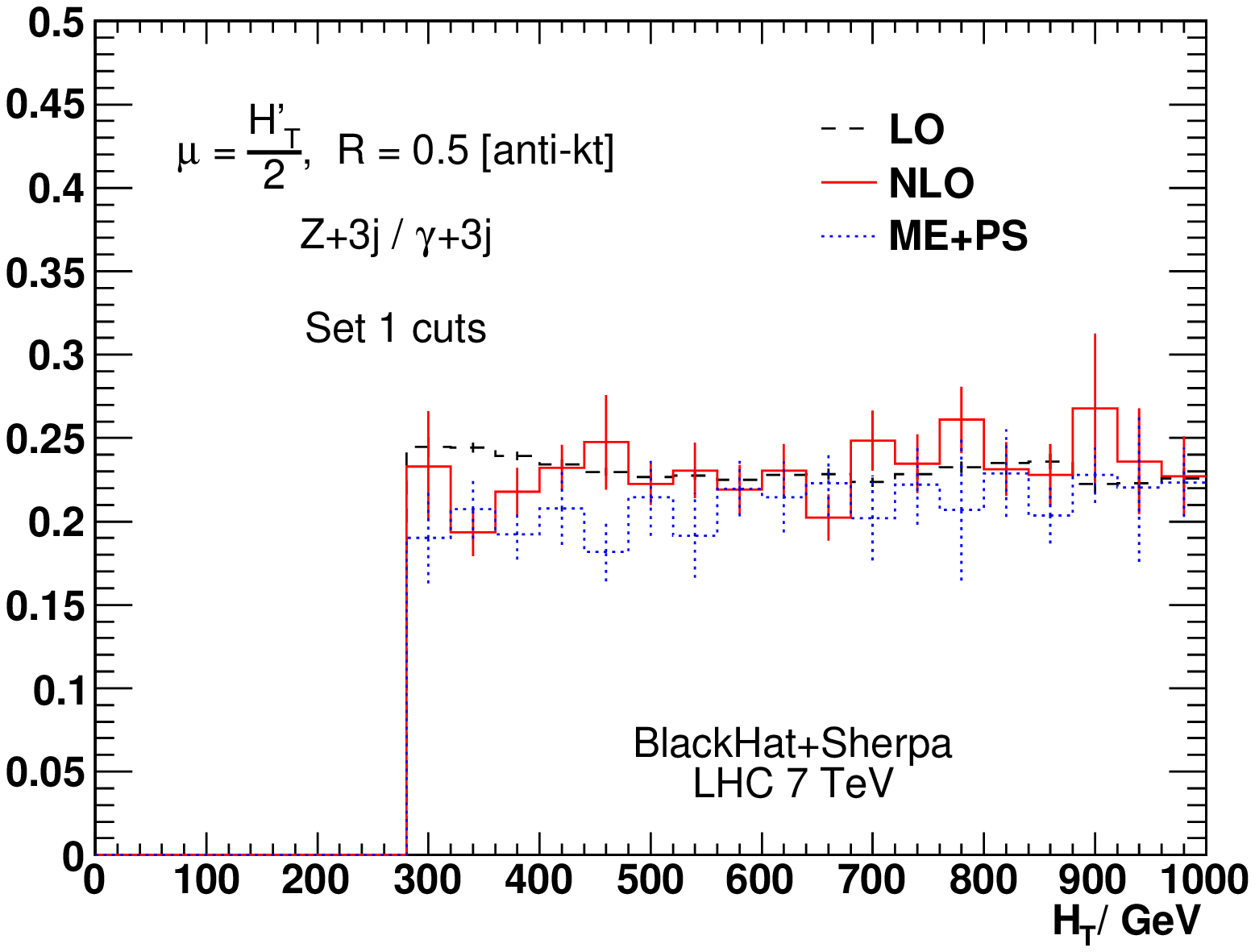}
    \includegraphics[clip,scale=0.40]{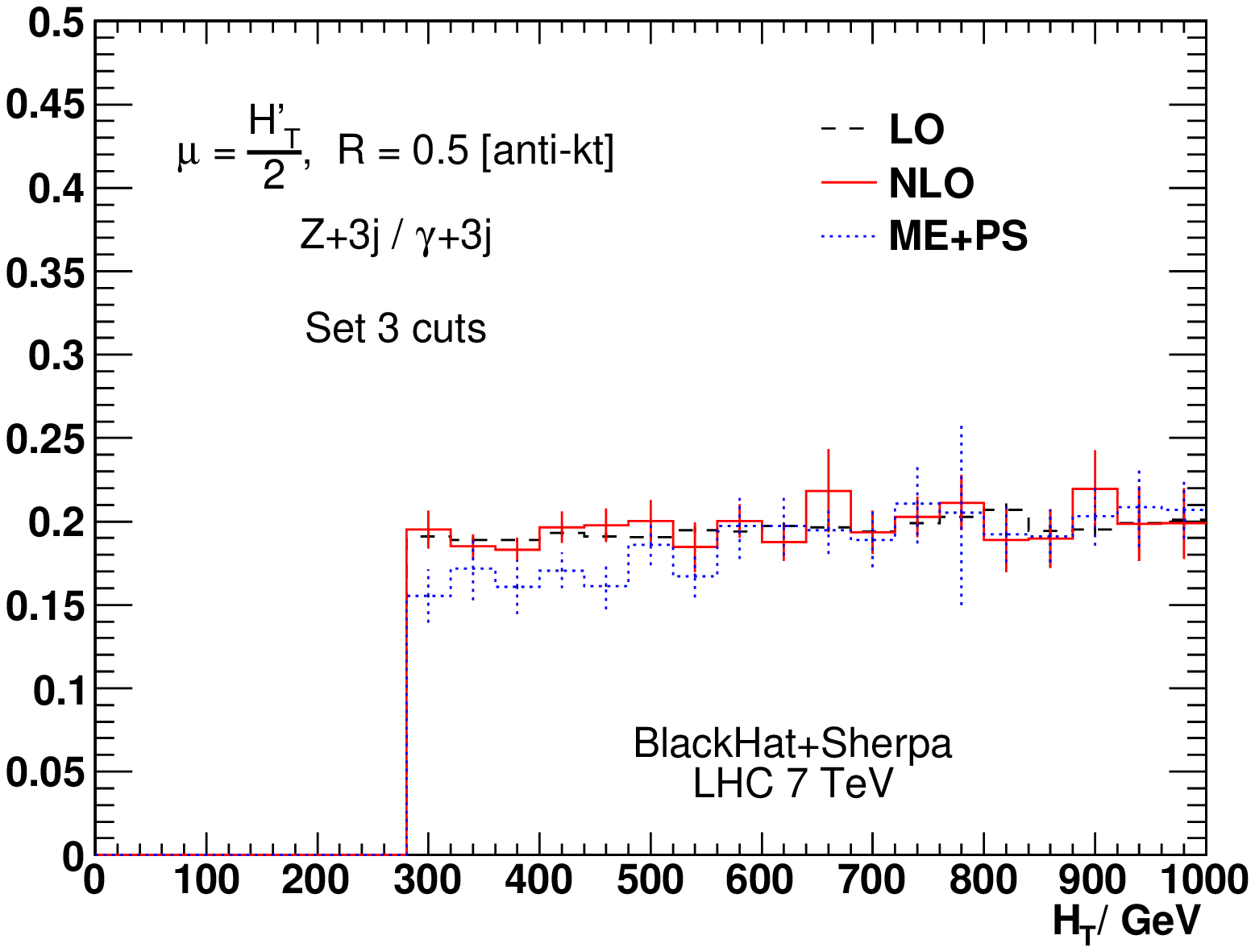}
    \includegraphics[clip,scale=0.40]{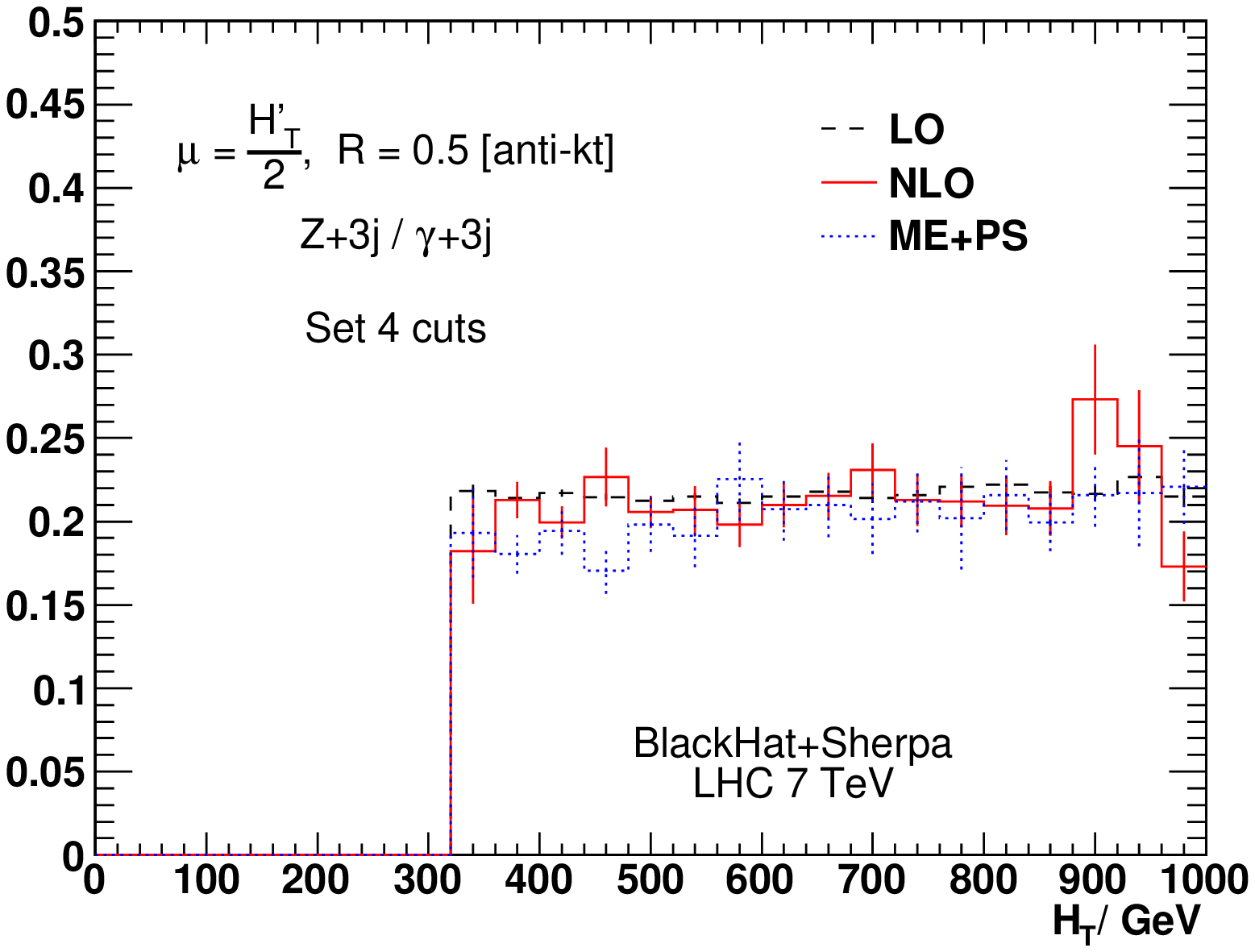}
    \includegraphics[clip,scale=0.40]{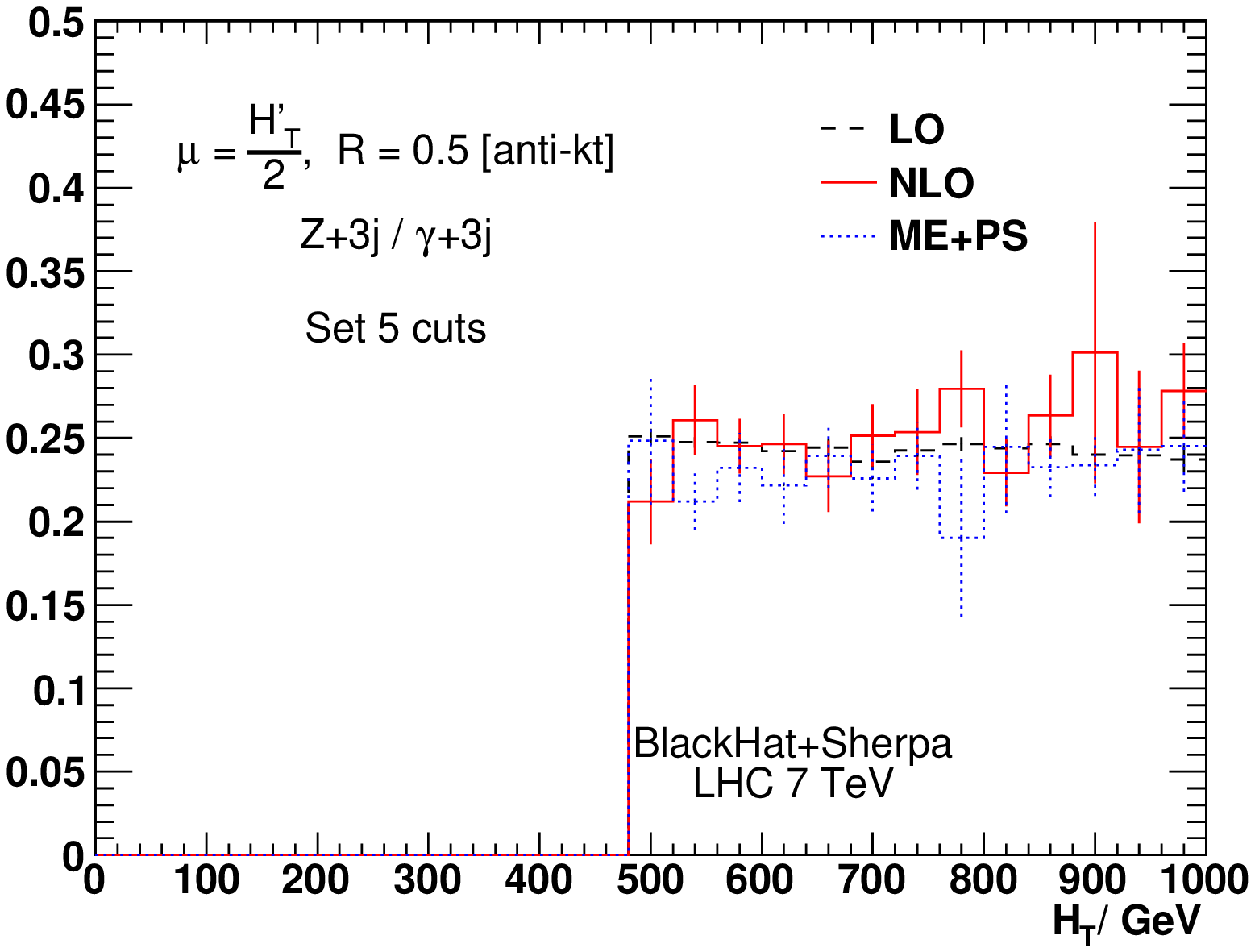}
    \includegraphics[clip,scale=0.40]{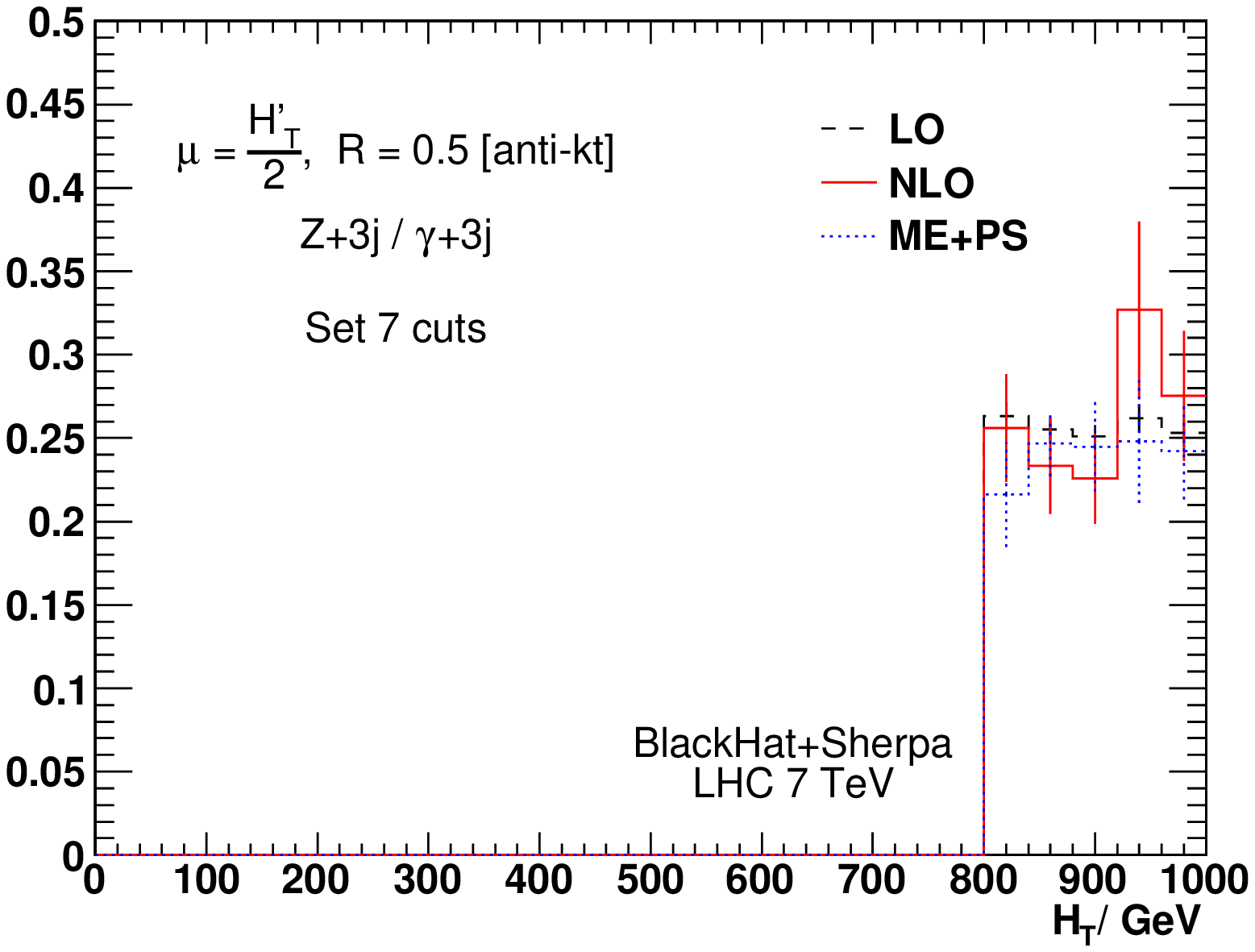}
  \end{center}
  \caption{\baselineskip 14pt The $\HTj$ distribution for the ratio of
    \Zjjj-jet to \gjjj-jet production for the different sets. We omit Sets 2
    and 6, as these plots are subsets of those for Sets 3 and 4
	respectively.}
  \label{fig:HT3}
\end{figure}

We have also computed various distributions.
\Fig{fig:HT3} shows the LO, NLO, and ME+PS predictions for
the $\HTj$ distributions for all control and signal sets except Set~2
and~6 (which are subsets of Sets 3 and 4 respectively).  The
NLO and ME+PS predictions for the
$Z$ to $\gamma$ ratio track each other well across the
whole range of $\HTj$, although in Sets~3 and~5 the shapes of the 
distributions are somewhat different.  (The total cross section in
each set is of course dominated by the lowest bins above the minimum
$\HTj$.)

As noted earlier, we have used a modified version of \SHERPA{}, based
on version 1.3.1, in order to ensure that the competition of
electroweak and QCD clusterings does not bias the $Z$ to $\gamma$
ratio.  In the unmodified version, the biasing effect is substantial
in the three-jet case, especially in the control regions (Sets 3 and 4).


\section{Uncertainty Estimates}
\label{UncertaintySection}

\subsection{QCD uncertainty}

\begin{table}[t]
  \begin{center}
\begin{tabular}{|c||l|l|l|l|l|l|l|}
\hline
\backslashbox{\phantom{XXl} Source}{\vspace{-8pt} Set} & 
\multicolumn{1}{c|}{1} & 
\multicolumn{1}{c|}{2} & 
\multicolumn{1}{c|}{3} & 
\multicolumn{1}{c|}{4} & 
\multicolumn{1}{c|}{5} & 
\multicolumn{1}{c|}{6} & 
\multicolumn{1}{c|}{7} \\
\hhline{|=#=|=|=|=|=|=|=|}
perturbative & ~~$0.09$~~ & ~~$0.03$~~ & ~~$0.10$~~ & ~~$0.07$~~ & ~~$0.05$~~ & ~~$0.02$~~ & ~~$0.04$~~ \\
PDF & ~~$0.02$~~ & ~~$0.03$~~ & ~~$0.02$~~ & ~~$0.02$~~ & ~~$0.03$~~ & ~~$0.04$~~ & ~~$0.05$~~ \\
photon-cone & ~~$0.01$~~ & ~~$0.01$~~ & ~~$0.01$~~ & ~~$0.01$~~ & ~~$0.01$~~ & ~~$0.01$~~ & ~~$0.01$~~ \\
\hline
total & ~~$0.09$~~ & ~~$0.04$~~ & ~~$0.10$~~ & ~~$0.08$~~ & ~~$0.06$~~ & ~~$0.04$~~ & ~~$0.06$~~ \\
\hline
\end{tabular}
\caption{Estimates of the fractional uncertainty remaining from QCD effects
for the \Zjjj-jet to \gjjj-jet 
ratios.  The ``perturbative'' uncertainty comes from comparing the
NLO ratio with the ME+PS one, as explained in the text.  The ``photon-cone''
  uncertainty is due to the estimated difference in predictions
using the standard and Frixione isolation cones.
 \label{AllSetsUncertaintiesH3} }
  \end{center}
\end{table}

As discussed in \sect{ZgammaStabilitySection}, the correlated scale
variation in the NLO calculation largely cancels in ratios, and does
not provide a suitable estimate of the remaining uncertainty due to
uncomputed higher-order corrections.  Instead, we use the NLO and
ME+PS ratios presented in \tab{AllSetsZgratio} to evaluate the
expected residual fractional uncertainty.  For each set, we do this by dividing
the absolute value of the difference between the two ratios by the NLO
ratio.  We add estimates of the PDF uncertainty, evaluated using
MSTW08 68\% error sets, and an estimate of the small uncertainty due to
using the Frixione cone in the theoretical calculation of photon cross
sections instead of the experimental fixed
cone~\cite{DrivingMissing}.
Although this somewhat overestimates the
uncertainty, we combined the PDF uncertainty in the numerator and
denominator in quadrature to arrive at the entries in \tab{AllSetsZgratio}.
We also combine the three uncertainties in the table
in quadrature to obtain the total uncertainty.  The estimates based on
inclusive \Vjjj-jet production are given in
\tab{AllSetsUncertaintiesH3}, and those based on \Vjj-jet production
are given in \tab{AllSetsUncertaintiesH2}.  One should be cautious in
taking estimates smaller than 10\% too literally, as the agreement
between NLO and ME+PS may not reflect all missing contributions
beyond that level.  These
uncertainty estimates should be taken symmetrically about the NLO
\Vjjj-jet ratio as a central value.  The overall uncertainty should be
at the 10\% level across all sets.  This is in agreement with the
estimate given in our earlier study~\cite{DrivingMissing}.  For at
least Sets 1--3, this theoretical uncertainty should be substantially
smaller than other experimental uncertainties in CMS's supersymmetry
limit~\cite{CMSSearch}.

\begin{table}[t]
  \begin{center}
\begin{tabular}{|c||c|c|c|c|c|c|c|}
\hline
\backslashbox{\phantom{XXl} Source}{\vspace{-8pt} Set} & 1 & 2 & 3 & 4 & 5 & 6 & 7 \\
\hhline{|=#=|=|=|=|=|=|=|}
perturbative & ~~$0.02$~~ & ~~$0.03$~~ & ~~$0.01$~~ & ~~$0.00$~~ & ~~$0.03$~~ & ~~$0.03$~~ & ~~$0.02$~~ \\
PDF & ~~$0.02$~~ & ~~$0.03$~~ & ~~$0.02$~~ & ~~$0.02$~~ & ~~$0.03$~~ & ~~$0.03$~~ & ~~$0.05$~~ \\
photon-cone & ~~$0.01$~~ & ~~$0.01$~~ & ~~$0.01$~~ & ~~$0.01$~~ & ~~$0.01$~~ & ~~$0.01$~~ & ~~$0.01$~~ \\
\hline
total & ~~$0.03$~~ & ~~$0.04$~~ & ~~$0.02$~~ & ~~$0.02$~~ & ~~$0.04$~~ & ~~$0.05$~~ & ~~$0.05$~~ \\
\hline
\end{tabular}
\caption{Uncertainty estimates for the \Zjj-jet to \gjj-jet 
ratios.  The labeling is as in \tab{AllSetsUncertaintiesH3}.
 \label{AllSetsUncertaintiesH2} }
  \end{center}
\end{table}

\subsection{A Partial Estimate of the Electroweak Uncertainty}

\begin{table}[t]
  \begin{center}
\begin{tabular}{|c||l|l|r@{}l|}
 \hline
Set & \Zjjj-jet (real EW) & \gjjj-jet (real EW) & \multicolumn{2}{l|}{{\Zjjj-jet/\gjjj-jet} (real EW)} \\
\hhline{|=#=|=|==|} 
1 & \dent$0.0250(0.013)$~~  & \dent$0.0162(0.017)$~~  & ~~~~~~~\dent$0$&$.0087(0.021)$~~ \\ \hline
2 & \dent$0.0169(0.016)$~~  & \dent$0.0151(0.016)$~~  & ~~\dent$0$&$.0017(0.022)$~~ \\ \hline
3 & \dent$0.0212(0.016)$~~  & \dent$0.0140(0.017)$~~  & ~~\dent$0$&$.0071(0.023)$~~ \\ \hline
4 & \dent$0.0221(0.012)$~~  & \dent$0.0152(0.014)$~~  & ~~\dent$0$&$.0068(0.019)$~~ \\ \hline
5 & \dent$0.0224(0.014)$~~  & \dent$0.0183(0.019)$~~  & ~~\dent$0$&$.0040(0.023)$~~ \\ \hline
6 & \dent$0.0139(0.016)$~~  & \dent$0.0179(0.019)$~~  & ~~\dent$-0$&$.0040(0.024)$~~ \\ \hline
7 & \dent$0.0208(0.018)$~~  & \dent$0.0221(0.023)$~~  & ~~\dent$-0$&$.0012(0.029)$~~ \\ \hline
\end{tabular}
\caption{
ME+PS predictions for fractional corrections
to $Z$ and $\gamma$ production in association with
three jets
  for the cuts of Sets 1--7 given in \sect{CutsSection}, where some of the jets may arise from the decay of an extra electroweak vector boson.
  The numbers in
  parentheses are Monte Carlo statistical errors. 
  \label{ElectroweakTables} }
  \end{center}
\end{table}

At very large values of MET and $\HTj$, the effects of electroweak
Sudakov logarithms are expected to become
important~\cite{ElectroweakLogsMaina, ElectroweakLogsKuhn}.  These
effects arise from virtual exchanges of electroweak bosons between
pairs of external partons or bosons that have large pair invariant
masses, well above the vector-boson masses.  In addition, one should
expect corrections due to the real emission of electroweak gauge
bosons from lower-jet multiplicity QCD processes, when the vector
bosons decay to jets. A complete calculation of these corrections is
beyond the scope of our present study.  We can however, make crude
estimates of the virtual corrections, along with a more reliable
estimate of the leading real-emission corrections using ME+PS matched
parton shower.  For the latter purpose, we used the same modified
version of SHERPA as in \sect{ZgammaStabilitySection}.

As a rough guide to the size of the electroweak virtual corrections,
we can use fig.~7 of ref.~\cite{ElectroweakLogsKuhn}.  This paper
studies electroweak-boson production accompanied by a lone jet.  For
$|\textrm{MET}| \sim 250$~GeV, the effects in the $Z$ to $\gamma$
ratio will be under 5\%; but for more aggressive cuts, $|\textrm{MET}|
\sim 500$~GeV as in Set 7, they could grow to 10\%.  For higher MET
cuts, the effects will grow beyond this value.  This estimate does not
take into account the additional jets, which for some subprocesses
increase the number of electroweak radiators, and which also increase
the partonic center-of-mass energy beyond that of the single-jet case.
We expect MET to be more important than $\HTj$ in determining the size
of the virtual electroweak corrections to the $Z$ to $\gamma$ ratio,
because the vector boson should have large invariant mass when paired
with another parton, in order to give a different correction factor
for a $Z$ boson versus a photon.

The larger number of electroweak radiators in the processes of
interest here may be expected to increase these effects somewhat, for
a fixed value of $|\textrm{MET}|$, although a significant fraction of
the cross section comes from subprocesses with a single quark line,
which have the same number of electroweak radiators as in the
calculation of ref.~\cite{ElectroweakLogsKuhn}.  
We expect these effects to increase the virtual contributions by 30\%
or so, say from a 10\% correction to a 13\% correction for
$|\textrm{MET}| \sim 500$~GeV.

The virtual corrections are of course partially canceled by real
emission of electroweak vector bosons~\cite{BaurEW}.  The latter
contribution can depend greatly on the observables and cuts.  We
expect the leading effect to be the emission of a $W$ or $Z$ boson
from a configuration with two fewer jets, with the hadronic decay of
the vector boson supplying the missing two jets.  (If the vector boson
is highly boosted, it might supply only a single merged jet.)  We
performed an ME+PS calculation, using \SHERPA{} to generate matched
matrix elements containing $WZ$, $ZZ$, $W\gamma$ and $Z\gamma$, along
with up to two additional partons.  The extra $W$ or $Z$ was then
decayed hadronically, and the decay products were treated on an equal
footing with the other jets in the event. In \tab{ElectroweakTables},
we present the contribution of this additional vector boson emission
to \Zjjj\ jets, to \gjjj\ jets, and to the ratio, as a fraction of the
basic calculation including only QCD emissions.  While the effects of
the electroweak real-emission contribution on the individual rates can
exceed 2\%, the corrections to the ratios are essentially negligible,
1\% or less across all sets.

Overall, we believe that the net electroweak corrections,
which are dominantly virtual, are likely to remain under 15\%, even for Set 7.
For even harder MET cuts, the effects may well become larger.
A precise calculation of the one-loop electroweak effects, even just
in leading or next-to-leading logarithmic accuracy, would clarify
these questions.


\section{Conclusions and Outlook}
\label{ConclusionSection}

In this paper, we have extended our previous
study~\cite{DrivingMissing} of the theoretical issues encountered
when using the measured \gp{}jets signal to
estimate the invisible $Z$+jets background in phase-space regions
selected by strong cuts suitable for supersymmetry searches.  In
particular, we have provided an estimate of the remaining theoretical
uncertainties in this translation.  In the previous study, we used the
\Zjj-jet to \gjj-jet ratio in two search regions, along with a control
region.  These regions correspond to the sets used by the CMS
collaboration in setting limits on supersymmetric
partners from the 2010 LHC data~\cite{CMSSearch}.
In this paper, we have extended the study
to an additional control region, and three new signal regions with
stronger cuts.  These new regions serve as a guide to searches with
harder cuts, as appropriate for larger data sets.
More importantly, we have added one more jet to the computation,
letting us study the \Zjjj-jet to \gjjj-jet ratio.

We computed the relevant differential cross sections and ratios to NLO
in QCD, and
estimated the remaining perturbative QCD uncertainty to be 10\% or
less by comparing with a parton-shower calculation, matched to LO matrix
elements (ME+PS), with the same number of jets.  As explained in the appendix,
we used a modified version of the \SHERPA{} matching algorithm
for this purpose.  We also studied uncertainties due to the parton
distribution functions, and found that they are 5\% or less in the
\Zjjx-jet to \gjjx-jet ratios.

We used the Frixione isolation criterion to compute the
prompt-photon cross sections.  In our previous study, we compared
isolated prompt-photon production with Frixione-cone isolation to that
with fixed-cone isolation at NLO, and found that the resulting shift should
be less than 1\% in the high-$\pt^\gamma$ region of interest. As part
of our present study, we have compared \gjjj-jet production using a
Frixione-cone and a fixed-cone in an ME+PS calculation; again we find a
difference of less than 1\% in the regions of interest.

Stronger cuts may also lead to larger QCD logarithms in
the \Vjjj-jet to \Vjj-jet ratios, which clouds our ability to rely on
purely perturbative predictions.  The smaller LO \Vjjjj-jet to
\Vjjj-jet ratios displayed in \tab{AllSets43ratioXS} suggest that this
is not crippling; also, we may expect the large QCD corrections to mostly
cancel in the \Zjjj-jet to \gjjj-jet ratio.  Experimenters could help
part these clouds, and restore full confidence in the applicability of
the perturbative uncertainty estimates in
tables~\ref{AllSetsUncertaintiesH3} and~\ref{AllSetsUncertaintiesH2},
by measuring the ratio of \gjjj-jet to \gjj-jet production in control
and search regions, and comparing these to the theoretical predictions
given in table~\ref{AllSets32ratioXS}.  We have not computed
potentially-significant Sudakov logarithms arising from virtual
electroweak corrections, but have given a crude estimate based on
ref.~\cite{ElectroweakLogsKuhn}. We have computed the leading
electroweak effects from emission of an additional $W$ or $Z$, and
find that these are 1\% or less in the $Z$ to $\gamma$ ratio, fairly
uniformly in all regions.

In summary, we find, across all search cuts, that the conversion
between photons and $Z$ bosons has less than a 10\% theoretical
uncertainty for events with either two or three associated jets.  This
is consistent with our previous findings~\cite{DrivingMissing}.  These
uncertainties are modest, and should make it possible for the photon
channel to provide a competitive determination of the Standard-Model
missing-$E_T$\,+\,jets background.  Furthermore, the NLO predictions
are remarkably robust under the addition of one jet, and should
ideally be used as the central value for experimental comparisons.

\section*{Acknowledgements}

We thank Anwar Bhatti, Mariarosaria D'Alfonso, Joe Incandela, Sue-Ann
Koay, Steven Lowette, Aneesh Manohar, Roberto Rossin, David Stuart and
Piet Verwilligen for extensive discussions.  This research was
supported by the National Science Foundation under Grant No. NSF
PHY05--51164, and by the US Department of Energy under contracts
DE--FG03--91ER40662, DE--AC02--76SF00515 and DE--FC02--94ER40818.
DAK's research is supported by the European Research Council under
Advanced Investigator Grant ERC--AdG--228301.  DM's work was supported
by the Research Executive Agency (REA) of the European Union under the
Grant Agreement number PITN-GA-2010-264564 (LHCPhenoNet).
The work of S.H. was partly supported by a grant from the US
LHC Theory Initiative through NSF contract PHY--0705682. This
research used resources of Academic Technology Services at UCLA and of
the National Energy Research Scientific Computing Center, which is
supported by the Office of Science of the U.S. Department of Energy
under Contract No. DE--AC02--05CH11231.


\appendix

\section{Modification to \SHERPA's ME+PS Algorithm}
\label{MEPSFix}

In this appendix we give a brief description of the ME+PS algorithm
used in \SHERPA, in order to explain how and why we modified the public
version 1.3.1.  The ME+PS method combines LO hard matrix elements
together with parton showers, which resum logarithmic corrections due
to gluon emission and parton splitting.  The parton shower we use in
\SHERPA{}~\cite{CSShower} is based on Catani-Seymour dipole
factorization~\cite{CS}. In contrast to earlier parton showers, the
procedure inherently respects QCD soft color coherence.  It
allows the unambiguous identification of a recoil partner for partons
that are shifted off their mass shell in the splitting process (the
``mother'' partons). This procedure eliminates one of the major sources of
uncertainty in earlier schemes for parton evolution.
We match the parton shower
to matrix elements containing up to four final-state partons,
and use 15~GeV for the merging cut.  (Further details may be found in
ref.~\cite{HoechePhoton}.) When matching a parton shower to LO matrix
elements using the CKKW algorithm~\cite{CKKWL}, one must cluster back
the matrix element configurations in order to define a parton-shower
starting condition.  In a shower that includes photons and electroweak
gauge bosons, the clustering should include the vectors as
well~\cite{KraussSchalicke}.

Two issues arise.  For the massive bosons, the reclustering is only
done approximately, due to missing helicity information.
This introduces a difference in the treatment of massive and massless
electroweak gauge bosons, which affects the $Z$ to $\gamma$ ratio
we are studying.  We will not study this issue here.
The other, and presumably more important, issue has to do
with the ordering of clusterings involving the vector-boson decay
products.  There is no parton-shower equivalent to the
$Z\rightarrow\nu\nub$ decay, of course, but in reducing a full \Zjjj-
or \Zjjjj-parton final state to a lower-multiplicity parton initiator,
the merging algorithm has to decide what to do with the neutrinos.
(For hadronic $Z$ decays there will be a full-fledged parton shower,
and for $Z$ decays to charged leptons additional QED radiation is possible.) 

Ideally, the parton-shower history should factorize into two independent
factors, one associated with $Z$ production, and the other with $Z$ decay.
However, in \SHERPA{} 1.3.1 the production and decay showers are interleaved.
In particular, the neutrinos from $Z$ decay are treated on an
equal footing with the partons when creating the clustering history
of an event (see sec.~4.4.2 of ref.~\cite{KraussSchalicke}).
The neutrinos are always produced with an invariant mass equal to the $Z$
mass.  Consequently, the treatment of radiation at scales below the $Z$
mass differs between \Zjn-jet and \gjn-jet production, affecting
precisely the ratio we wish to compute.  In more detail, the
clusterings involving an electroweak object compete with the QCD clusterings, 
and the competition goes differently in the $Z$ and $\gamma$ cases.
This creates a bias in the weighting of parton-shower initiators.
The photon splitting amplitudes are simply color-stripped versions of
the QCD ones, so the competition in the case of the photon gives the
correct result.  The bias shows up in ME+PS calculations of
\Zjn-jet production.

The effect turns out to be substantial in the high-$\pt$ regions
of interest in our study. Our modification to version 1.3.1
of \SHERPA{} is simply to force the neutrino pair to cluster to a $Z$
at the first CKKW step, and then to remove the associated scale from the
cluster history. This modification guarantees that we treat the $Z$ and
photon identically in the ME+PS algorithm.


\end{document}